\documentclass[twoside]{article}

\usepackage{PRIMEarxiv}

\usepackage[utf8]{inputenc} 
\usepackage[T1]{fontenc}    
\usepackage{hyperref}       
\usepackage{url}            
\usepackage{booktabs}       
\usepackage{amsfonts}       
\usepackage{nicefrac}       
\usepackage{microtype}      
\usepackage{lipsum}
\usepackage[ruled,vlined]{algorithm2e}
\usepackage{fancyhdr}       
\usepackage{graphicx}       
\usepackage[labelformat=simple]{subcaption}

\DeclareCaptionLabelFormat{subcaptionlabel}{\normalfont(\textbf{#2}\normalfont)}
\graphicspath{{media/}}     

\title{Learning Complex Spatial Behaviours in ABM: An Experimental Observational Study.
\thanks{\textit{\underline{Citation}}: 
\textbf{Authors. Title. Pages.... DOI:000000/11111.}} 
}

\author{
  Sedar Olmez \\
  School of Geography \\
  University of Leeds \\
  Seminary St, Woodhouse, Leeds LS2 9JT\\
  The Alan Turing Institute \\
  2QR, John Dodson House, 96 Euston Rd, London NW1 2DB\\
  \texttt{\ gysol@leeds.ac.uk} \\
   \And
  Dan Birks \\
  School of Law \\
  University of Leeds \\
  Belle Vue Rd, Woodhouse, Leeds LS2 9JT\\
  The Alan Turing Institute \\
  2QR, John Dodson House, 96 Euston Rd, London NW1 2DB\\
  \texttt{D.Birks@leeds.ac.uk} \\
  \And
  Alison Heppenstall \\
  School of Social \& Political Sciences \\
  University of Glasgow \\
  University of Glasgow Adam Smith Building Bute Gardens Glasgow, G12 8RT\\
  The Alan Turing Institute \\
  2QR, John Dodson House, 96 Euston Rd, London NW1 2DB\\
  \texttt{alison.heppenstall@glasgow.ac.uk} \\
}

\begin{document}
\maketitle

\begin{abstract}
Capturing and simulating intelligent adaptive behaviours within spatially explicit individual-based models remains an ongoing challenge for researchers. While an ever-increasing abundance of real-world behavioural data are collected, few approaches exist that can quantify and formalise key individual behaviours and how they change over space and time. Consequently, commonly used agent decision-making frameworks, such as event-condition-action rules, are often required to focus only on a narrow range of behaviours. We argue that these behavioural frameworks often do not reflect real-world scenarios and fail to capture how behaviours can develop in response to stimuli. There has been an increased interest in Machine Learning methods and their potential to simulate intelligent adaptive behaviours in recent years. One method that is beginning to gain traction in this area is Reinforcement Learning (RL). This paper explores how RL can be applied to create emergent agent behaviours using a simple predator-prey Agent-Based Model (ABM). Running a series of simulations, we demonstrate that agents trained using the novel Proximal Policy Optimisation (PPO) algorithm behave in ways that exhibit properties of real-world intelligent adaptive behaviours, such as hiding, evading and foraging.
\end{abstract}


\section{Introduction}
In this research, intelligent adaptive behaviour is defined as \textit{agents using their knowledge about the environment to make decisions that allow them to adapt to new and novel situations}. Understanding how these behaviours change over space and time is vital to understanding the processes that shape complex systems and how they change. Social systems, for example, contain dynamic processes that evolve over space and time. \cite{Osullivan2012Agent-basedIt} discusses the changes that occur at the neighbourhood level and describes the social processes that are created by these changes.  According to \cite{Osullivan2012Agent-basedIt}, social systems are driven by potentially infinite individual-level decisions. \cite{Batty2013TheCities}, develops this view by suggesting that social systems are comprised of individuals linked by networks with information transmitted between them. Any attempt to simulate individual decision-making and behaviour must capture social processes at the individual level.

An ABM (a type of individual-based model) is a computational model used for simulating the actions and interactions of autonomous agents \cite{Ji2017MathematicalSystems}. ABM has grown in popularity over the past decade; this widespread adoption is observed by the applications of these models in various domains, for example, economics, sociology, criminology, geography \cite{Malleson2010CrimeBurglary, Olner2015AnEmergence, Crooks2014AnCholera, Heppenstall2006UsingMarkets, Heppenstall2012Agent-basedSystems}. Most recently, some scholars have considered how ABM might be applied to support our understanding of the Covid-19 pandemic and societal responses to it \cite{Squazzoni2020ComputationalAction, Cuevas2020AnFacilities}.

While the upsurge in new data sources, including footfall cameras, pollution monitors and mobile navigation, provides novel insights into behaviour and movements of individuals \cite{Luo2008Agent-basedSimulation, Dawson2011AnManagement, Benenson2008PARKAGENT:City}, identifying individual-level behaviours to embed within ABM remains a challenge. However, efforts have been directed towards constructing frameworks for handling behaviour in ABMs \cite{Balke2014HowSurvey, DeAngelis2019Decision-makingProspectus, Groeneveld2017TheoreticalReview}. 

Almost all ABMs implement some aspect of decision-making \cite{Malleson2013UsingRisk, Epstein1999Agent-basedScience, Crooks2014AnCholera, Kangur2017AnVehicles, Olmez2021ExploringApproach}. However, the frameworks applied to these models vary in their purpose and complexity. A common drawback of ABMs is that pre-determined decision rules constrain the decision-making process; these are often based on the analysis of historical data. Such models can only replicate a narrow set of behaviours often constrained to a specific period through their decision-making. \cite{DeAngelis2019Decision-makingProspectus} points out that many factors must be considered when decisions are made, such as the environment, the rewards and penalties the environment provides, the complex internal state of the organism and its incomplete knowledge of the environment.

\cite{Balke2014HowSurvey} states that the process of modelling complex decision-making has many levels of complexity, including cognitive (such as reactive, deliberative, neurologically inspired agents), learning (the learning agents undergo, including neural networks), social (including communication and social networks). By contrast to existing agent decision-making frameworks, this research will explore how agents can learn to exhibit intelligent adaptive behaviours that vary across space and time. To achieve this, this paper presents the results of efforts to imbue agents within a simple illustrative predator-prey model with behaviours derived through reinforcement learning (RL) - a well-known machine learning technique.

The goal of RL algorithms is to learn how to apply an action in a given situation; RL does this by mapping situations (states) to actions through trial-and-error and uses an objective function to measure how well a set of actions are in maximising the rewards. Thus, the agent learns behaviour and is not told explicitly what action to take  \cite{Sutton2018ReinforcementEdition,Wooldridge2020TheMachines}. A real-world example of such behavioral learning is a dog and its owner; the owner rewards the dog with treats when it behaves and tells it off if it displays unwanted reactions such as barking or biting. Over time, the dog is conditioned to learn to behave appropriately (as defined by the owner).

RL has previously been applied within ABM research, \cite{Lopes2018IntelligentLearning} for instance, applied RL to simulate adaptive drone behaviour; the goal was to learn how to cover unfamiliar terrain to get to a target. \cite{Sert2020SegregationModeling} further developed Schelling's model, a well-known ABM which explores segregation dynamics \cite{Schelling1969ModelsSegregation}, utilising an RL decision-making framework to capture novel insights, such as evidencing that segregated areas are more likely to host older people than diverse areas, which attract younger ones. \cite{Spatharis2019CollaborativeManagement} harnessed the group-learning ability of RL (multi-agent systems where agents learn collectively as a group) to develop collaborative management dynamics of air traffic control. Similarly, \cite{Nawrocki2018AdaptiveLearning} applied RL to develop policies for service management of cloud computing infrastructure. The agent's goal is to gain experience in determining the optimal place in which a given task should be executed. Most similar to this study, \cite{baker2019emergent} conduct an exploratory study that demonstrates how PPO can be used to develop intelligent agent behaviour. In their study \cite{baker2019emergent} demonstrate how static environmental changes (i.e. new platforms, barriers) impact the behaviour of interacting agents. While not explored by \cite{baker2019emergent}, RL may also offer means to explore how dynamic changes to the model impact behaviour - likely key to 'adaptive' behaviour - and something that more traditional ABM decision-making frameworks would struggle to capture. Within the model in this paper, this idea is extended by exploring how dynamic stimuli unbeknown to agents during training, effect the decisions agents develop post-training. Furthermore, another unexplored area by the aforementioned study is the length of training time. There is a consensus among some scholars in RL that training models for longer lead to agents that are better at completing tasks \cite{Kim2019EffectsStudy, Juliani2018Unity:Agents}, we test this idea in a relatively small scale.

While these studies have demonstrated that RL is capable of exploring intelligent behaviour, the implementation of RL presents a number of significant challenges. Firstly, there are currently no systematic methods available to validate the behaviours that emerge from RL, while by contrast condition-action behavioural rules are relatively easily inspected as the modeller knows which action is being applied as this is implemented as a rule in the model. Secondly, logistically speaking RL models are both more complex to implement and the computational complexity associated with their application is many orders of magnitude greater than traditional decision making frameworks, making RL appealing to only those with access to high-performance computing facilities. The impact of this is that fewer models are developed and fewer applications are explored.
\label{para1.8}

This research will evaluate the utility of a relatively new RL algorithm, namely proximal-policy optimisation \cite{Schulman2017ProximalAlgorithms} for simulating intelligent adaptive behaviours and the subsequent variability of spatial patterns that emerge from these behaviours. This algorithm was chosen as it has been shown to outperform the majority of RL algorithms it was benchmarked against (sub-section \ref{PPO_sub}) \cite{Schulman2017ProximalAlgorithms}.

This research is not attempting to describe empirically validated ecologically verified behaviours of predator-prey interactions. However, the research does develop a simple illustrative model containing two interacting entities that can easily be tested and examined.

The objective of this research is to demonstrate the usability of RL in developing intelligent adaptive behaviours in an illustrative agent-based model and subsequently, to interpret these emerging behaviours qualitatively and quantitatively. To achieve this objective, we devise two experiments, where (1) looks at the impact training length has on task efficiency, answering the question: does learning for a longer duration lead to behaviours with better outcomes than those trained for a shorter period? In experiment (2), three model scenarios are devised, these are: prey agents are trained without the predator, and the predator is not present post-training (Scenario 1), prey agents are trained with the predator, and predator is present post-training (Scenario 2); lastly, the prey agents are trained without the predator, while the predator is introduced post-training (Scenario 3). The question to be answered is: how do agents adapt to the presence of an unknown stimulus? Does this have an adverse effect on task efficiency? Given the two experiments, we aim to compile outputs from these experiment scenarios and (1) assess the quantitative outputs by comparing task efficiency across the different experiment scenarios. (2) analyse the individual behaviours from recorded simulations to interpret intelligent adaptive behaviours.
\label{objectives_paragraph}

The research objective is achieved by: developing a simple ABM containing two types of agents (prey and predator) in the Unity software platform using the ml-agents software package \cite{Juliani2018Unity:Agents}. Training this model under several experimental conditions using PPO \cite{Schulman2017ProximalAlgorithms}, and subsequently examining the outcomes of the trained models both quantitatively and qualitatively. Finally, a framework is devised to record, analyse and interpret real-time behaviours agents portray during the simulation runs.

\subsection{Simulating Behaviour using Reinforcement Learning}
Many ABMs within the literature study behaviour within a particular domain, for example, ecological economics, animal movement or crime \cite{Heckbert2010Agent-basedEconomics, Tang2010Agent-basedReview, Birks2012GenerativeTheory, Matthews2007Agent-basedApplications}. Typically, these models use frameworks that handle reactive behaviours or contain agents that adopt frameworks in line with bounded rationality \cite{Selten2001What,Arthur1994InductiveProblem,Manson2006BoundedPrograms}, such as belief-based methods \cite{Wooldridge2002ReasoningAgents}. This can make it difficult to see how reflective of the real world these models can be. \cite{Dawid2011Agent-basedDesign} describe the issues that ABMs for economic policy encounter. Once they deviate from perfectly rational agents, they encounter many degrees of freedom on what to assume for the behaviour of an agent.

Intelligent adaptive behaviours are challenging to develop using current agent-based decision-making frameworks as these behaviours can be inconsistent or unexpected (irrational) \cite{Kennedy2012ModellingModels}. Some argue that these behaviours are constrained in the following way: a behavioural entity should be able to perceive its environment, should have a knowledge database and several degrees of freedom of action on the environment \cite{Newell1994UnifiedCognition}.

The goal of RL is to allow agents to experience their immediate environment, gather information from this environment using sensors, and over time learn which combination of decisions (policies) can lead to the desired outcome. This training phase (sub-section \ref{training_section}) allows a neural network (Figure \ref{fig:ANN_architecture}) to capture action, state, outcome data, which agents can use to infer decisions post-training. The main concern of this research is the behaviours that subsequently emerge during the agents' activities post-training and how these behaviours can be quantified \cite{Wooldridge2020TheMachines,Kaelbling1996ReinforcementSurvey,Sutton2018ReinforcementEdition,Busoniu2010Multi-agentOverview}.

In their research on applying behavioural frameworks in competitive game scenarios to study strategy equilibria, \cite{Erev1998PredictingEquilibria} outlines the main theories inspired by behavioural psychology that RL adopts; these are The Law of Effect \cite{Thorndike1927TheEffect}: decisions that have led to positive outcomes in the past, are more likely to be repeated in the future. The Power Law of Practice \cite{Blackburn1936TheCurves}: Learning curves initially tend to be steep and then flatter over time. After numerous experiments comparing RL against belief-based models, researchers found that RL is more responsive, allowing it to adapt to other agents' changing behaviour, improving the models' predictive power. Furthermore, when comparing higher-rationality decision making, such as belief-based learning to lower-rationality RL, the former does not have an advantage over the latter on the datasets tested \cite{Erev1998PredictingEquilibria}.

The strengths of RL include knowledge retention (collecting observational sensory data and devising policies using these data to achieve long-term goals) of agents and knowledge sharing between agents \cite{Clouse1996LearningAgent}. The drawbacks of RL agents include the exponential growth of the discrete state-action space in the number of state and action variables and the exploration-exploitation trade-off. This trade-off requires RL algorithms to balance the exploitation of the agent's current knowledge and exploratory information-gathering actions taken to improve that knowledge \cite{Busoniu2010Multi-agentOverview}.

Some RL applications in ABMs focus on energy expenditure, economic forecasting, and game theory. \cite{Thapa2005AgentCircumstances} developed an agent-based decision support system for a patient diagnosis. The authors apply an RL algorithm as it provides approximation methods to make trade-offs between accuracy and speed. As a result, many cases were solved in a shorter period. Similarly, \cite{Jalalimanesh2017Simulation-basedLearning} developed a simulation for optimising the radiography process using ABMs and RL. The proposed approach contains two steps; the first is the ABM, which simulates different radiotherapy scenarios. The second step involves Q-learning \cite{Watkins1992Q-learning}, a popular RL algorithm used to optimise the radiation dose and fractionation scheme. Finally, the learning algorithm parameters are fine-tuned until the optimal treatment plans are achieved to cure tumours with minimal side effects.

Generally, RL applications in modelling can be split into two categories. In category (1), we have studies that adopt RL in their simulation to identify the optimal solution to a particular problem \cite{Lopes2018IntelligentLearning,Spatharis2019CollaborativeManagement,Nawrocki2018AdaptiveLearning,Thapa2005AgentCircumstances,Jalalimanesh2017Simulation-basedLearning}. In category (2), we have studies that use RL as a decision-making framework to investigate a known phenomenon and observe new insights from this phenomenon, that may have not previously been identified due to constrained decision-making frameworks \cite{Sert2020SegregationModeling,Erev1998PredictingEquilibria,SegismundoS.Izquierdo2008ReinforcementDilemmas,FLACHE2016StochasticCooperation,Zschache2017TheLearning,Zschache2016MeliorationGames}. This research does not fall into either of the categories; instead, it aims to apply RL to agents such that intelligent adaptive behaviours organically grow across multiple training conditions, as would be the case in the real-world. Capturing and interpreting these behaviours could, in future, allow models to represent realistic behaviours better.

Another contribution made by this research is to introduce specialised software, namely game engines, to the academic modelling community; in the following paragraph, the rich features available in Unity that can help modellers develop more realistic models are described. 

Unity is a 3D development platform that consists of a rendering and physics engine. Unity has received
widespread adoption in several industries, including gaming, automotive, and film \cite{Juliani2018Unity:Agents}. Modern games development software such as Unity are powerful tools for simulating complex interactions between agents with varying physiological/mental capacities. Thus, games development software are perfectly poised to provide solutions for the foreseeable future of AI research \cite{Juliani2018Unity:Agents}. Furthermore, advances in RL have primarily been driven by the ability of neural networks to process large amounts of visual, auditory and text-based data \cite{Lecun2015DeepLearning}. Many of the applied tasks researchers are interested in solving with AI involve not only rich sensory information but a rich control scheme in which agents can interact with their dynamic environments in complex ways \cite{Bicchi2000RoboticReview, baker2019emergent}.

In Section \ref{section2}, the PPO RL algorithm is described, including how it works, its features and relative strengths and weaknesses. Section \ref{section3} outlines the ABM, the types of agents, environment, rewards and penalties that agents yield. Section \ref{section4} defines the steps taken during training of the RL model; this includes the parameters used for training, formal definition of these parameters and subsequently, the results from the training process. Section \ref{section5} details the simulation results; these results are quantitatively analysed using various data science techniques. The individual behaviours from these experiments are qualitatively described; a systematic approach was developed to ascribe visually inspected behaviours to determine intelligent adaptive behaviours. Finally, Section \ref{section6} discusses the research outcome, what was learnt from the research and how the research would be useful in future applications conducted by ABM researchers.

\section{Implementing Reinforcement Learning Algorithms in Unity}
\label{section2}
This section describes the novel PPO RL algorithm \cite{Schulman2017ProximalAlgorithms,Juliani2018Unity:Agents}. The algorithm is illustrated by outlining the main components, including the underlying formulae and pseudocode.

\subsection{Proximal Policy Optimisation (PPO)}
\label{PPO_sub}
PPO is a policy gradient method; these are a type of RL methods that depend on optimising parameterised policies concerning the expected return (long-term cumulative reward) by gradient descent \cite{Sutton2018ReinforcementEdition}.

The PPO algorithm used in this research was developed by OpenAI researchers \cite{Schulman2017ProximalAlgorithms}. These algorithms alternate between sampling data by interacting with the environment and optimising a proxy objective function using the stochastic gradient descent algorithm \cite{Ruder2016AnAlgorithms}. The developers of PPO argue that given the recent advancements made in RL algorithms that adopt neural network function approximators. There are still areas that could be improved, such as making these algorithms scalable to larger models (even more so given the COVID-19 pandemic and the need for large scale models that simulate population behaviours), parallel computation applications and solving multiple problems without the need for hyper-parameter tuning \cite{Schulman2017ProximalAlgorithms}.

During testing, \cite{Schulman2017ProximalAlgorithms} compared the PPO algorithm against algorithms that are known to perform well at solving continuous control problems. The algorithms compared were: Trust Region Policy Optimisation (TRPO) \cite{Schulman2015TrustOptimization}, Cross-Entropy Method (CEM) \cite{Szita2006LearningMethod}, Advantage Actor-Critic (A2C) \cite{Mnih2016AsynchronousLearning}, A2C with Trust Region \cite{Wang2016SampleReplay}. During these tests, PPO outperformed the algorithms mentioned above in almost all instances of continuous control scenarios.

\begin{equation}\label{2.1}
L^{CPI}(\theta)=\hat{\mathbb{E}}_{t}\left [ \frac{\pi_{\theta}(a_{t}|s_{t})}{\pi_{\theta old}(a_{t}|s_{t})}\hat{A}_{t} \right]=\hat{\mathbb{E}}_{t}\left [ r_{t}(\theta)\hat{A}_{t} \right ]
\end{equation}

In the above Formula \ref{2.1}, CPI stands for "conservative policy iteration" \cite{Kakade2002ApproximatelyLearning}. Without a constraint, maximisation of $L^{CPI}$ would lead to an extremely large policy update; therefore, the objective function needs to be modified to penalise changes to the policy that shift $r_{t}(\theta)$ away from 1. Subsequently, the following Formula \ref{2.2} was developed \cite{Schulman2017ProximalAlgorithms}.
\par
The previous variant of the PPO algorithm detailed in \cite{Schulman2017ProximalAlgorithms} used an adaptive Kullback-Leibler divergence (a measure of how one probability distribution is different from a second, reference probability distribution) \cite{Kullback1951OnSufficiency} penalty to control the change of policy at each iteration. The newly updated variant of the PPO algorithm adopts a different objective function (a method to measure the quality of any solution to a problem) proposed by \cite{Schulman2017ProximalAlgorithms} which can be found below (Formula \ref{2.2}).

\begin{equation}\label{2.2}
L^{CLIP} (\theta) = \hat{\mathbb{E}}_t[min(r_t(\theta)\hat{A}_t,clip(r_t(\theta),1-\varepsilon, 1+\varepsilon)\hat{A}_t)]
\end{equation}

Where $\theta$ is the policy parameter, $\pi$ is the policy, $a$ and $s$ are action and state respectively, $ \hat{\mathbb{E}}_t$ is the empirical expectations over time steps. $r_t$ is the ratio of the probability under the new and old policies, respectively. $\hat{A}_t$ is the estimated advantage at time $t$. $\varepsilon$ is a hyperparameter, usually between 0 and 1; the hyperparameter value is used to control the learning process. As described by \cite{Schulman2017ProximalAlgorithms}, the first term inside the min is $L^{CPI}$ (Formula \ref{2.1}). The second term, $clip(r_t(\theta),1-\varepsilon, 1+\varepsilon)\hat{A}_t)$, adjusts the surrogate objective by clipping the probability ratio, which eliminates the incentive for moving $r_t$ outside of the period $[1 - \varepsilon, 1 + \varepsilon]$. Finally, the minimum of the clipped and unclipped objective is taken, so the ultimate objective is a lower bound (also known as a pessimistic bound) on the unclipped objective. Given this system, the change in probability ratio is ignored if the objective improves; conversely, it is only included when it makes the objective worse.

The PPO Actor-Critic algorithm outlined by \cite{Schulman2017ProximalAlgorithms} is defined below in pseudocode (see Algorithm \ref{algorithm1}). For each iteration, every agent adopts an initial policy (a set of action/state combinations) $\pi_{\theta_{old}}$ (as utilised in Formula \ref{2.1}) in the environment for $T$ time steps. The advantage estimates $\hat{A}_t$ (as utilised in Formulas \ref{2.1} and \ref{2.2}) are calculated for each time step. The algorithm then constructs the surrogate loss given the policy parameter $\theta$ on these $NT$ time steps of data and optimises it with a Minibatch Stochastic Gradient Descent (this is a variation of gradient descent that splits training data into small batches that are used to calculate model error and update model coefficients) for $K$ epochs.

\begin{algorithm}[!h]
\SetAlgoLined
 initialization\;
 \For{iteration=1,2...}{
  \For{actor=1,2,...,N}{
    Run policy $\pi_{\theta_{old}}$ in environment for T timesteps\\
    Compute advantage estimates $\hat{A}_1$,...,$\hat{A}_T$\
  }
    Optimise surrogate $L$ wrt $\theta$, with $K$ epochs and minibatch size $M \leq NT$\\
    $\theta_{old}\leftarrow\theta$
 }
\caption{PPO, Actor-Critic Style}
\label{algorithm1}
\end{algorithm}

The ability to deploy PPO in Unity as a decision-making framework for agents with relative ease and its performance against other learning algorithms outlined previously, made it the top contender of algorithms to adopt in this research. Similarly, being able to solve multiple problems with varying complexities without the need to tune training parameters made it suitable for the experimental conditions we aim to conduct \cite{Juliani2018Unity:Agents, Schulman2017ProximalAlgorithms, baker2019emergent}.

In ml-agents \cite{Juliani2018Unity:Agents} PPO uses an artificial neural network (ANN) to approximate the ideal function that maps an agent's observations to the best action an agent can take in a given state (Figure \ref{fig:ANN_architecture}).

Robotics researchers adopted PPO to develop a Mobile robot navigation application whereby robots learn to navigate a terrain without any knowledge of the map \cite{Zeng2018LearningNavigation}. Similarly, researchers adopted PPO to simulate a multi-agent environment between two groups of agents. The actor-critic variant of PPO, where a policy network produces an action distribution and the critic network predicts the discounted future returns \cite{Sutton2018ReinforcementEdition} produced some valuable results, proving that PPO can be used to simulate complex behaviours \cite{baker2019emergent}. A common drawback from these specific studies is the spatial patterns that emerge from environmental changes that were not explored in detail. We know from the literature that individual behaviours can be different and evolve through time \cite{Sutton2018ReinforcementEdition,Juliani2018Unity:Agents,baker2019emergent,Sert2020SegregationModeling}; however, what do these individual behaviours look like? What are the environmental implications of these individual behaviours? These are the questions this study is interested in.

The drawbacks of PPO are: acquiring good results via Policy Gradient methods is demanding because they are sensitive to the choice of step size - too small, and progress is unbearably slow; too large, and the signal is overwhelmed by the noise \cite{Schulman2017ProximalAlgorithms}. As model complexity increases, solving these problems using RL can become computationally intensive; therefore, in some cases, high-performance computing clusters may be required to adopt PPO in research. Lastly, trained agents may be unable to adapt to changes in the environment; this is commonly referred to as overfitting in machine learning. One mechanism to alleviate this and train more efficient agents is to expose agents to these changes during training \cite{Juliani2018Unity:Agents}. As a model becomes more complex and training environments become more dissimilar to test scenarios, overfitting becomes more likely. It is recommended to use separate training/test scenarios (with varying levels of stochasticity) while ensuring some similarities \cite{Zhang2018ALearning}. This study proposes a simplistic model with only two agent types and a static environment, with relatively short training time than other large-scale models \cite{Lopes2018IntelligentLearning, Spatharis2019CollaborativeManagement, Nawrocki2018AdaptiveLearning}. Therefore, overfitting is less likely to occur and could be identifiable when individual behaviours are quantitatively and qualitatively interpreted.

\section{Model Description}
\label{section3}
The model contains two agents: a predator agent to catch the prey agents; and prey agents that avoid the predator and forage points. The model simulates agents interacting in a simple three-dimensional environment that contains physical barriers that block the vision and movement of all agents. Through several experiments, this research explores how agents that adopt PPO devise intelligent adaptive behaviours, given the environmental surroundings they find themselves in.

Predator-prey models are not new; they have been previously used to simulate the interactions of wildlife in ecology \cite{Zhdankin2010SimpleModel, Colon2015BifurcationInteractions, Hawick2008DefensiveModel}. As mentioned in the Introduction section, the predator-prey scenario is a simple scenario to model. Thus, this domain was chosen to promote explainability and reproducibility.

In this research, interest is centred around learned behaviours that emerge given the environmental factors the agents find themselves in. For example, barriers were added to the environment to test if prey agents utilise them in hiding from the predator. Prey agents have no inherent defence mechanism to deploy against the predator. The behaviours we aim to observe have been identified in past literature, especially those in fish biology research of predator-prey behaviours. \cite{doi:10.1111/j.1095-8649.1978.tb03432.x} describe the interactions among two types of fish; they describe the primary defence for prey fish as "detection of pursuit". This behavioural strategy includes "fish visually tracking the predator's movement, and when it swims within their line of sight, they then get away from it at the nearest opportunity". Given the behaviours observed in these predator-prey scenarios, it is expected that the following behaviours are observed in the model: 

\begin{itemize}
    \item prey evade the predator.
    \item prey forage rewards.
    \item prey use the environment to hide from the predator.
\end{itemize}

This research is interested in training agents to learn and enact simple to more complex spatial behaviours. Some of these complex behaviours are outlined above. Thus, the environment has been designed to motivate the above behaviours and allow prey to utilise these behaviours in novel situations.

\subsection{Purpose}
The ABM was developed in Unity using the ml-agents software package \cite{Juliani2018Unity:Agents}, the model allows the modeller to simulate hypothetical activities of prey and predator agents in 3D space. 

\subsection{Agents}
Prey agents are rewarded for foraging positive points placed on the environment and penalised for foraging negative points and being caught by the predator. The predator agent follows any prey that falls within its field of view.

The prey agents adopt the PPO RL framework for learning, while the predator agent's decision-making framework is a set of condition-action rules such as IF prey-within-view THEN chase-prey. The predator moves around the environment randomly and tries to catch prey agents while the prey agents learn how to adapt to this. Once the predator physically touches the prey agent, the prey agent is caught, leading to a penalty. The agents were created this way to reduce the impact on computational demand as having two types of agents, both applying RL, would be computationally expensive.

For the predator agent parameters, Table \ref{tab:Predator_parameters} and Table \ref{tab:Prey_parameters} for the prey agent parameters. The values for each agent type are relative to that agent's size and mass. For both agents, an appropriate view radius was chosen relative to the environment and the agent's physical features, preventing one from physically being superior. 

In the model, prey agents share the same characteristics. This ensures no prey agent has abilities that can make it superior to others, e.g. speed or vision - as this would arguably introduce additional complexity in interpreting model outcomes during the two experiments (as described in Section \ref{para1.8}). The environment randomly distributes positive points (which prey collect) and negative points (which prey should avoid). These point objects are used as a training indicator; if the reward increases, agents are learning (Figure \ref{fig:training-1}).

\begin{figure}[!h]
\centering
\includegraphics[width=0.4\textwidth]{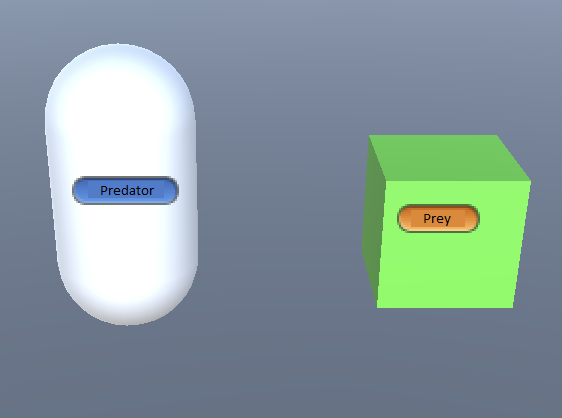}
\caption{The predator agent (left) and a prey agent (right) in the Environment.}
\label{fig:predator_and_prey_agents}
\end{figure}

\begin{figure}[!h]
\centering
\includegraphics[width=0.8\textwidth]{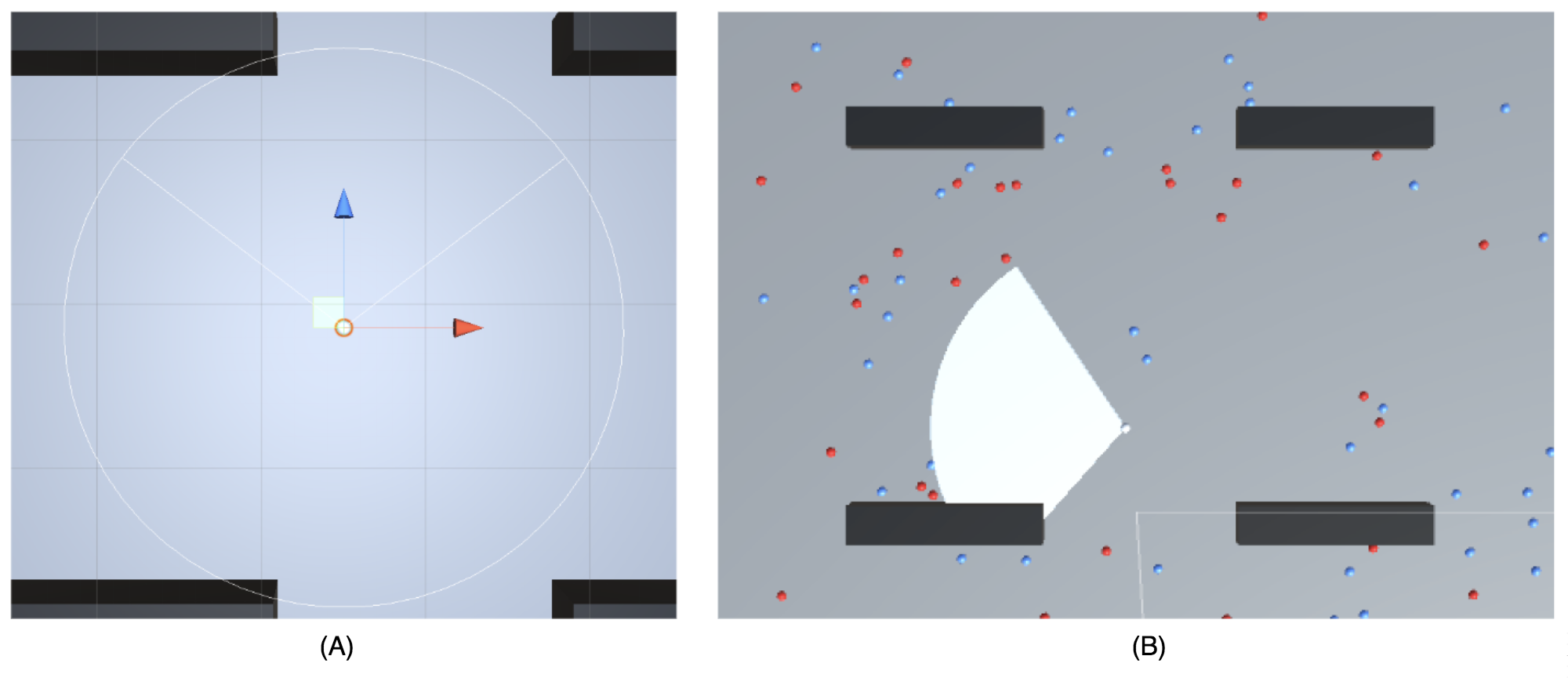}
\caption{The area in which the predator can see (A). The model environment and the vision cone of the predator looking for prey agents (B).}
\label{fig:predator_agent}
\end{figure}

\begin{figure}[!h]
        \centering
        \includegraphics[width=0.8\textwidth]{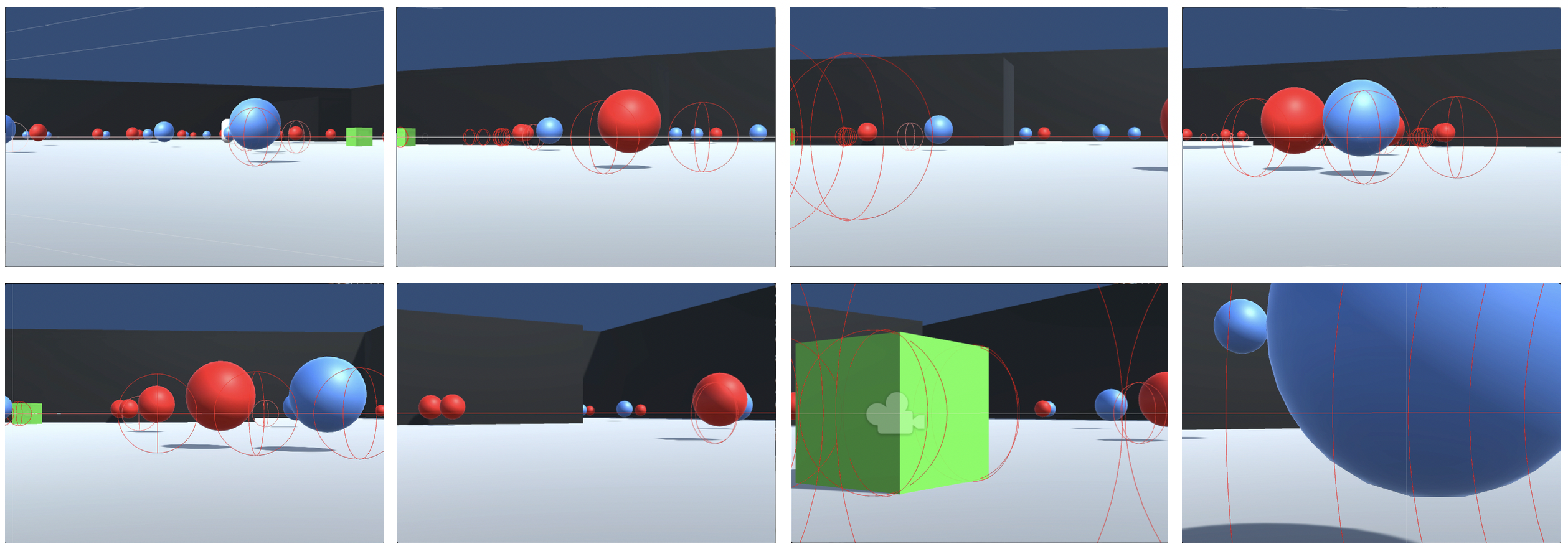}
        \caption{First person view from Prey agent's perspective.}
        \label{fig:appendix_1}
\end{figure}

The physical representation of both agents can be seen in Figure \ref{fig:predator_and_prey_agents}. The predator field of view and viewing angle allows it to identify objects (Figure \ref{fig:predator_agent}). Prey agents move on the surface of the environment and can perceive the world through ray-cast sensors collecting observations. A first-person observation can be found in Figure \ref{fig:appendix_1}.

The predator moves randomly around the environment until a prey agent falls within its vision cone; this is the patrolling phase. The predator makes its movement unpredictable; thus, prey agents can be trained for all circumstances. As mentioned earlier, the condition-action rules for the predator are;

\begin{itemize}
    \item Chase the prey if the prey is within view.
    \item While the simulation is running, move randomly on the surface of the environment.
\end{itemize}

\subsection{Environment}
The environment includes the following Unity components;

\begin{itemize}
    \item Plane - a 3D flat surface area for agents to stand on.
    \item Wall - a 3D object that acts as a barrier stopping agents from falling off the plane.
    \item Camera - A camera pointing at the environment and agents.
    \item Directional light - A light ray pointing at the environment with soft shadows helps the observer see the environment.
\end{itemize}

The environment provides the prey agents with enough information to allow them to learn intelligent adaptive behaviours. If barriers were not present, the prey agents could not learn how to hide. Similarly, if the predator does not exist, there is no motivation for prey agents to learn how to evade capture. Each element of the environment has several customisable properties (Table \ref{tab:Environment_parameters}) and can be changed depending on specific requirements.

\begin{figure}[!h]
\centering
\includegraphics[width=0.4\textwidth]{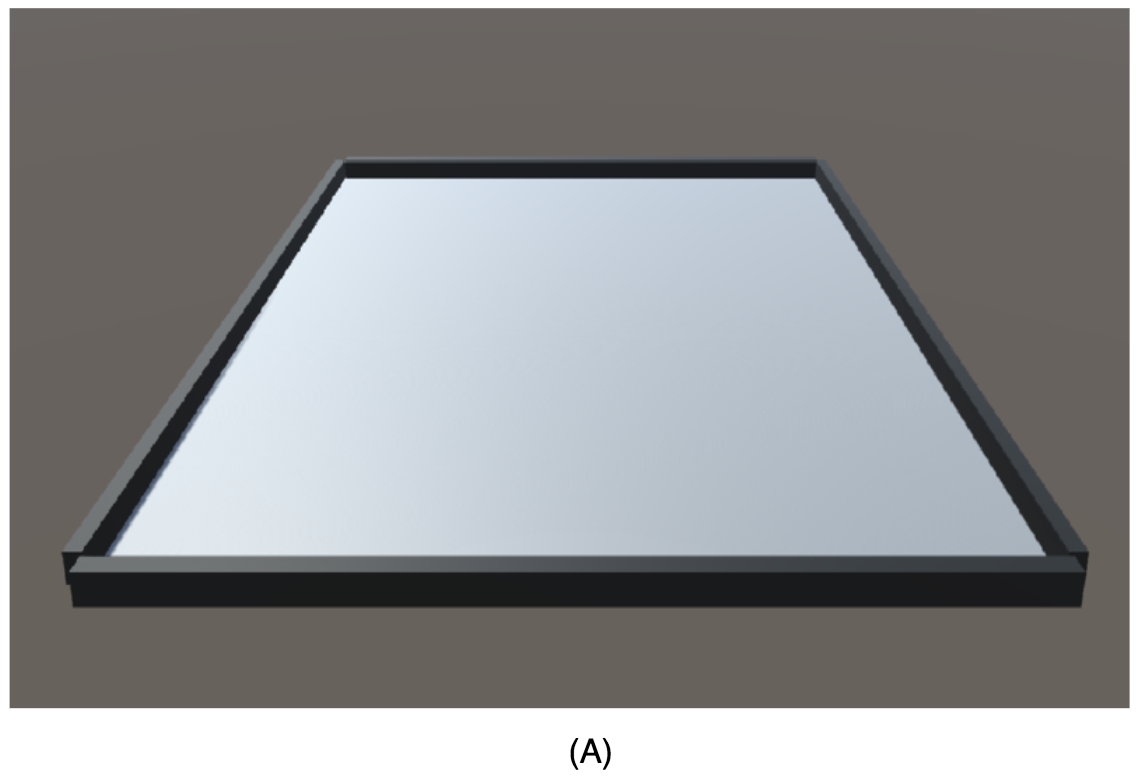}
\includegraphics[width=0.4\textwidth]{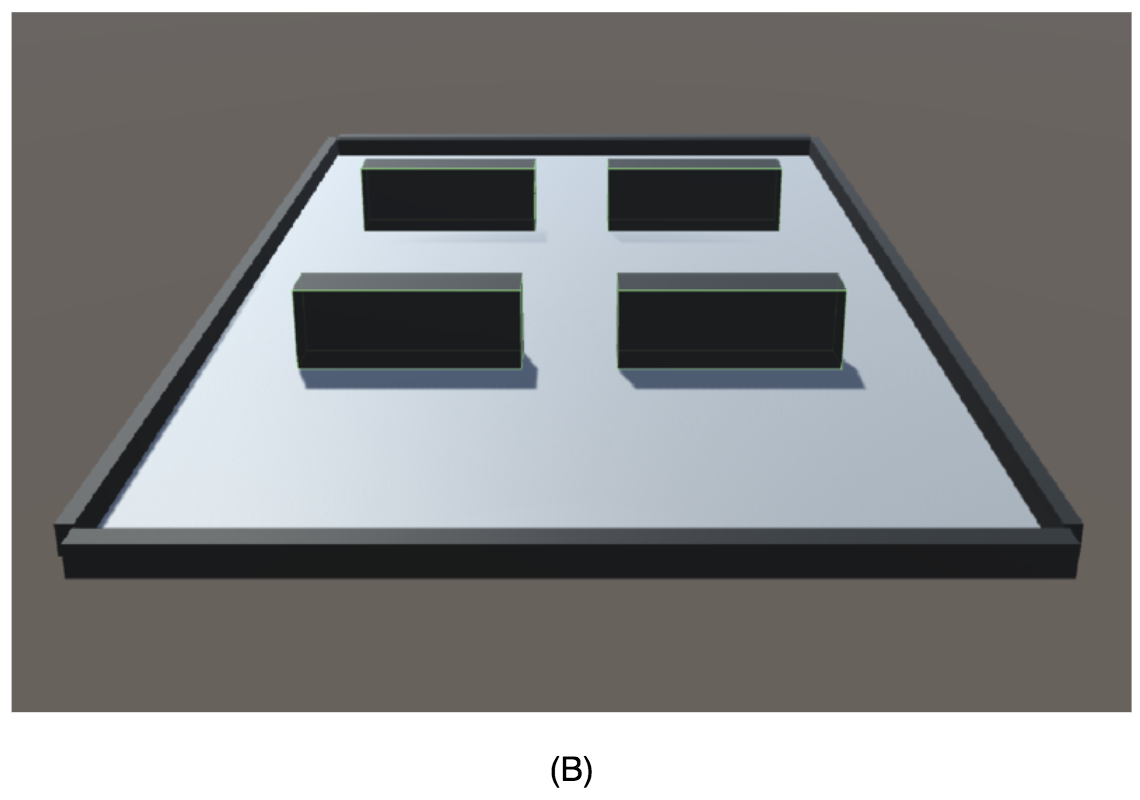}
\caption{The initial state of the environment (A). The environment scene once parameters from Table \ref{tab:Environment_parameters} are applied (B).}
\label{fig:Environment-Illustration-1}
\end{figure}

The parameters described in Table \ref{tab:Environment_parameters}, if implemented, would produce the 3D environment scene in Figure \ref{fig:Environment-Illustration-1} (B).

\begin{figure}[!h]
\centering
\includegraphics[width=0.6\textwidth]{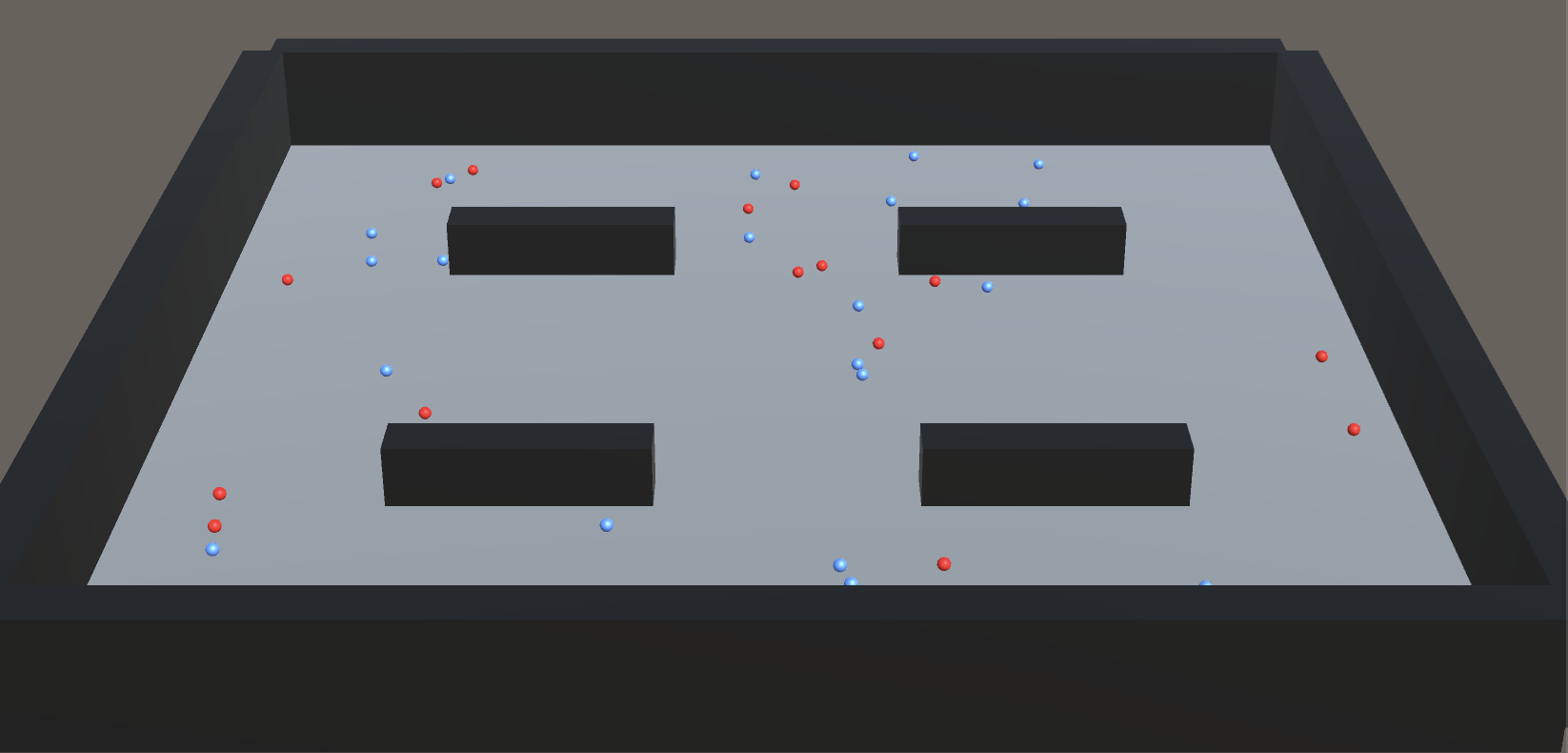}
\caption{Environment with positive point objects (blue spheres) and negative point objects (red spheres).}
\label{fig:Environment-Illustration-2}
\end{figure}

The positive and negative points are randomly distributed on the environment surface so that the prey agents can forage them (Figure \ref{fig:Environment-Illustration-2}). If the points are collected, they re-appear at a random location within the environment. For every point collected, the prey agent is either rewarded or penalised.

\section{Training Process}
\label{section4}
To successfully train an RL algorithm, training parameters are selected to ensure the performance of learning processes and quality of generated motions \cite{Kim2019EffectsStudy, Juliani2018Unity:Agents}. An RL model is performing well if the cumulative reward is increasing during training \cite{poole2010artificial}. To conduct the experiments outlined in Section \ref{objectives_paragraph}, three neural network models will be trained using parameters that coincide with the experiment objectives, the differences between experiments are the training length and presence of a novel stimulus Table \ref{tab:Hyper-parameters-three-scenarios}.

\subsection{Training Parameters}\label{training_section}
\begin{figure}[!h]
\centering
\includegraphics[width=0.80\textwidth]{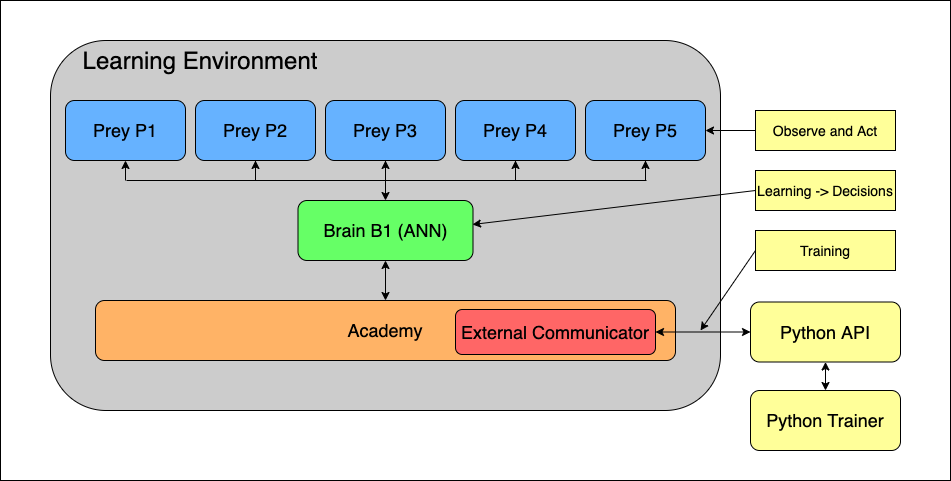}
\caption{A block diagram of ml-agents}
\label{fig:ml-agents-diagram}
\end{figure}

The ml-agents package \cite{Juliani2018Unity:Agents} simplifies the training process of artificial agents in Unity (Figure \ref{fig:ml-agents-diagram}). The Learning Environment component contains the Unity scene, which includes the environment agents can act, observe and learn from. The "brain" component takes the observed data from agents (known as Vector Observations) and is trained using the Academy (Table \ref{tab:Hyper-parameters-three-scenarios}). The Academy connects the brain to the python trainer, where the artificial neural network training commences. Once the training process ends (Figure \ref{fig:training-1}), the output neural network is attached to the agents post-training, and thus, the agents can infer decisions from the trained model (Tables \ref{tab:exp1_summary} and \ref{tab:exp2_summary}).

\begin{figure}[!h]
        \centering
        \includegraphics[width=0.8\textwidth]{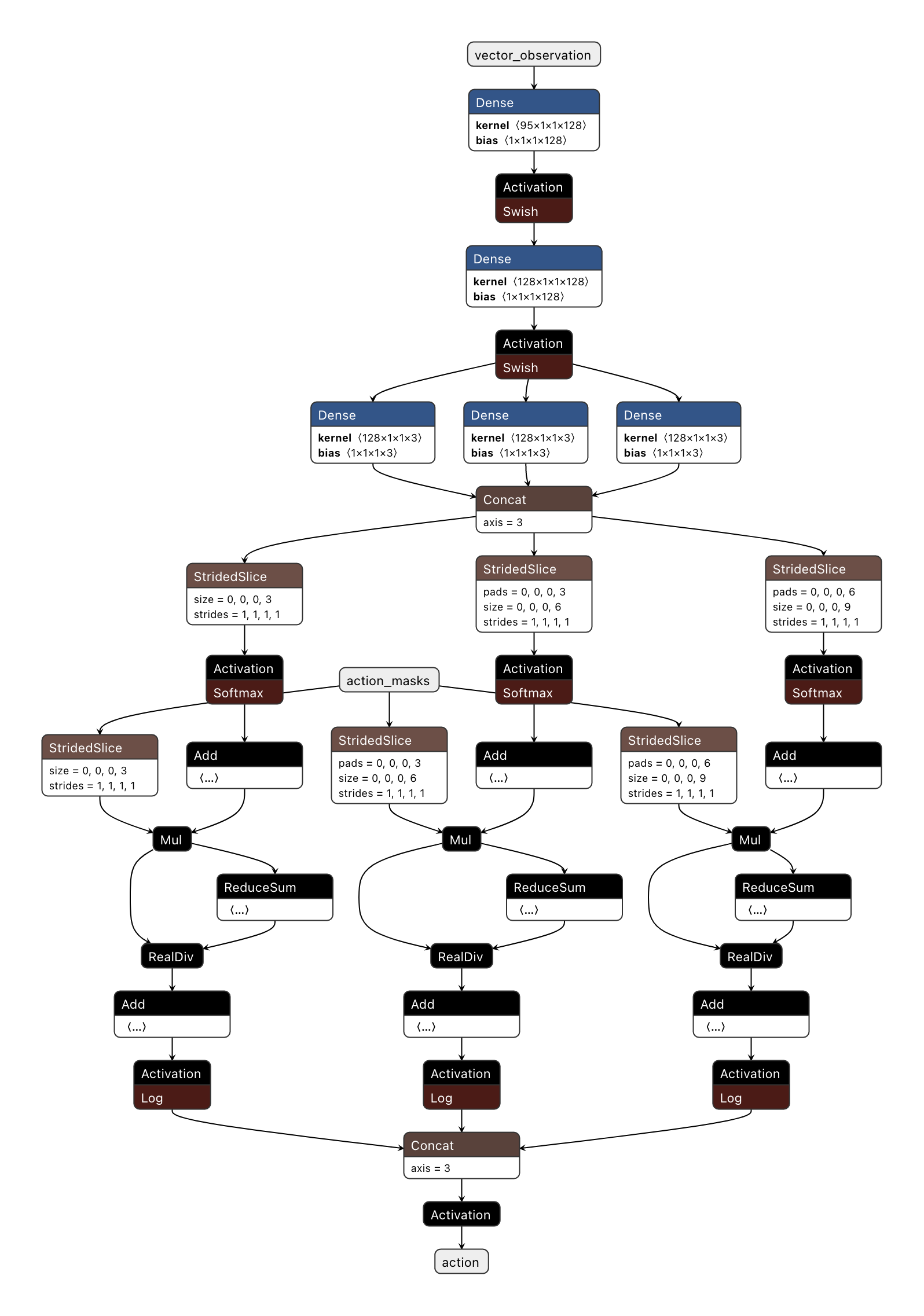}
        \caption{Artificial Neural Network Architecture.}
        \label{fig:ANN_architecture}
\end{figure}

In this research, the default training parameters provided by ml-agents \cite{Juliani2018Unity:Agents} were used for training; ml-agents researchers recommended these parameters as suitable for the majority of tested environments (Table \ref{tab:Hyper-parameters-three-scenarios}). Furthermore, these parameters were within the recommended ranges (Table \ref{tab:Hyper-parameters-formal-definition}). These parameters led to successful learning sessions (Figure \ref{fig:training-1}). The training process outputted three fully trained artificial neural network models (ANNs). For the model architecture, refer to Figure \ref{fig:ANN_architecture}.

\begin{table}[!h]
\centering
\resizebox{\textwidth}{!}{\begin{tabular}{p{3cm}p{5cm}p{5cm}p{5cm}}
\toprule
Parameter & Scenario 1 (\textbf{w/predator}) & Scenario 2 (\textbf{w/predator}) & Scenario 3 (\textbf{wo/predator}) \\
\midrule
Trainer & PPO & PPO & PPO \\
Batch\_size & 1024 & 1024 & 1024 \\
$\beta$ & 1.0e-2 & 1.0e-2 & 1.0e-2\\
Buffer\_size & 10240 & 10240 & 10240\\
$\epsilon$ & 0.2 & 0.2 & 0.2 \\
Hidden\_units & 128 & 128 & 128 \\
GAE $\lambda$ & 0.95 & 0.95 & 0.95 \\
Learning\_rate & 3.0e-4 & 3.0e-4 & 3.0e-4 \\
Learning\_rate\_schedule & Linear & Linear & Linear \\
Max\_steps & \textbf{580000} & \textbf{1.0e6} & \textbf{1.0e6} \\
Memory\_size & 256 & 256 & 256 \\
Normalize & false & false & false \\
Num\_epoch & 3 & 3 & 3 \\
Num\_layers & 2 & 2 & 2 \\
Time\_horizon & 64 & 64 & 64 \\
Sequence\_length & 64 & 64 & 64 \\
Summary\_freq & 10000 & 10000 & 10000 \\
Use\_recurrent & false & false & false \\
Reward\_signals & extrinsic: strength: 1.0, $\gamma$: 0.99 & extrinsic: strength: 1.0, $\gamma$: 0.99 & extrinsic: strength: 1.0, $\gamma$: 0.99  \\
\bottomrule
\end{tabular}}
\caption{PPO training parameters for all three scenarios.}
\label{tab:Hyper-parameters-three-scenarios}    
\end{table}

In Table \ref{tab:Hyper-parameters-three-scenarios}, the only differences between the models are the max\_steps variable for scenario one, which is 580,000-time steps, and in scenario three, the predator was not present during training. While training, the algorithm only started to successfully learn how to increase cumulative rewards at 100,000-time steps. This was identified as the lower bound. To then identify the upper bound and the point at which cumulative rewards stabilise, the model was trained for longer, and 1,000,000 time steps were identified as the point in which rewards were no longer being gained (scenarios two and three in Table \ref{tab:Hyper-parameters-three-scenarios}). The reason behind this process is to produce models that coincide with the two experiments outlined earlier. For other models or research questions, the training max\_steps would be different given the model complexity.

\cite{6151508} describe the adaptability of an RL agent by "making a major or minor change to the system after it has learnt the optimal policy and analyse the learning algorithm adapt to these changes". To test this theory, model training configuration three (scenario three in Table \ref{tab:Hyper-parameters-three-scenarios}) was conceived. Configurations two and three were trained for the same time. The former was trained with the predator, while the latter was trained without the predator.

\newpage
\subsection{Training Results}
\begin{figure}[!h]
\centering
\includegraphics[width=0.99\textwidth]{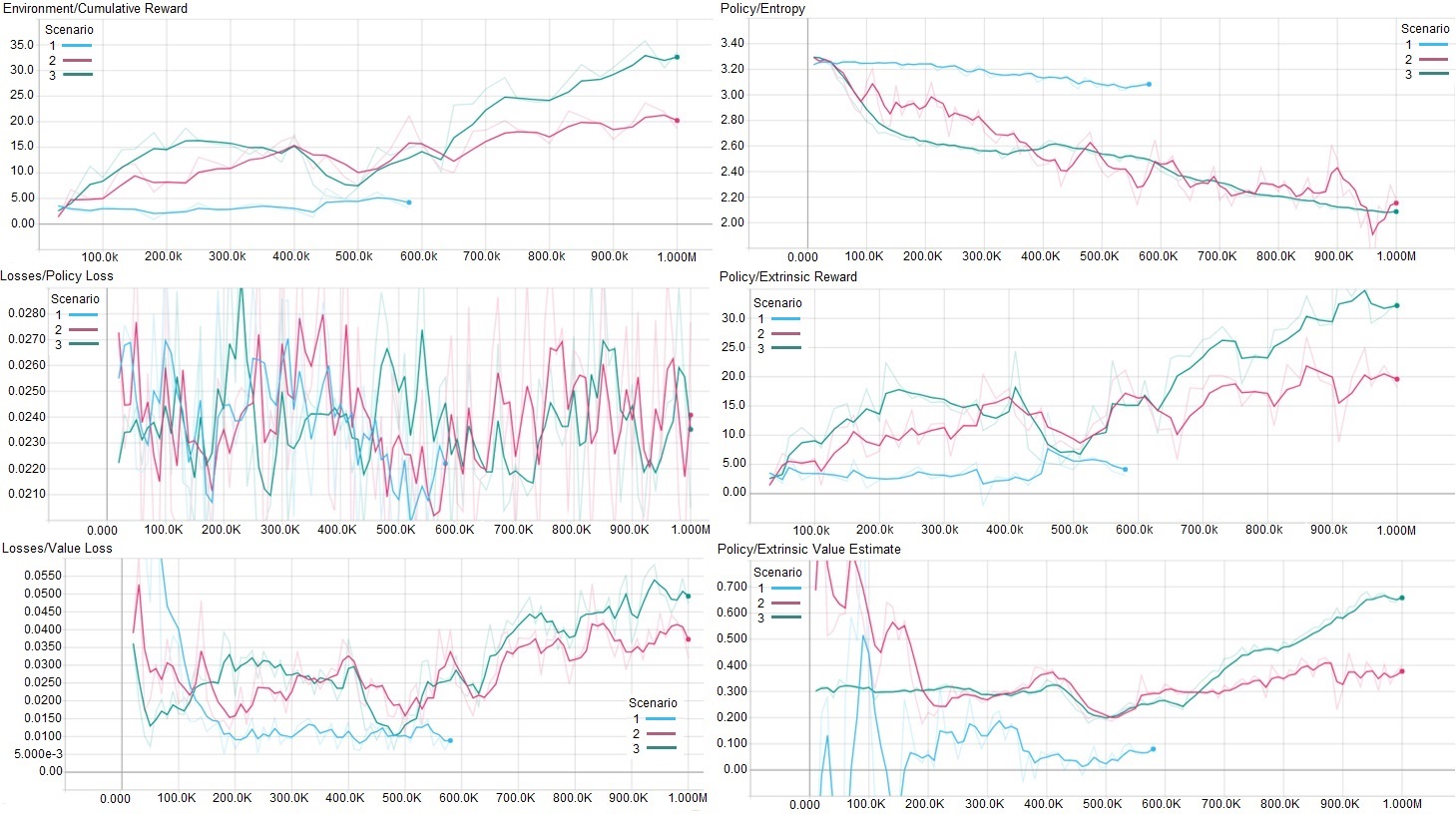}
\caption{Six graphs from the PPO training process each line corresponds to a scenario the model applied during training, outlined in Table \ref{tab:Hyper-parameters-three-scenarios}. x-axis: number of time-steps in training, y-axis: value.}
\label{fig:training-1}
\end{figure}

Figure \ref{fig:training-1} highlights the results from the training process outlined earlier in Table \ref{tab:Hyper-parameters-three-scenarios}.

Descriptions of the results presented in Figure \ref{fig:training-1} can be found below:
\begin{itemize}
\item Cumulative reward (Figure \ref{fig:training-1}, top left) - the mean cumulative episode reward over all agents should increase during a successful training session.
\item Policy loss (Figure \ref{fig:training-1}, middle left) - the mean magnitude of the policy loss function. Correlates to how much the policy (process for deciding actions) is changing. The magnitude of this should decrease during a successful training session.
\item Value loss (Figure \ref{fig:training-1}, bottom left) - the mean loss of the value function update. Correlates to how well the model predicts the value of each state, should increase while the agent is learning, and then decrease once the reward stabilises.
\item Entropy (Figure \ref{fig:training-1}, top right) - represents how random the decisions of the model are. Should slowly decrease during a successful training process. If it decreases too quickly, the $\beta$ parameter should increase.
\item Extrinsic reward (Figure \ref{fig:training-1}, middle right) - this corresponds to the mean cumulative reward received from the environment per episode.
\item Extrinsic value estimate (Figure \ref{fig:training-1}, bottom right) - the mean value estimate for all states visited by the agent. Should increase during a successful training session.
\end{itemize}

In Figure \ref{fig:training-1}, we observe that the training parameters for scenario three were the most successful compared to scenarios one and two. The only difference between scenarios two and three was the predator. Thus, these training outcomes show that the predator's presence impacts how the prey agents behave regarding rewards; when the predator is not present, the prey agents can forage more rewards. Furthermore, agents' policies frequently change over time, meaning prey agents identify good policies more frequently. The Policy Extrinsic Reward results suggest that for all three scenarios, prey agents tend to increase their rewards over time; this also means they are more likely to reduce their penalties, including predator avoidance. Ultimately, the results from these data show that all three training scenarios and model setup was successful, i.e. prey agents were designed to successfully learn policies from their immediate environment, which they can apply post-training. Lastly, scenarios one and two led to different training outcomes (the only difference was the max\_steps parameter); the reason for this is the stochasticity of the model; for each epoch, prey, points and predator are randomly distributed.

\section{Simulation Results}
\label{section5}
Before examining the behaviours learnt by the prey agents operating under PPO, several experiments are developed. The training phase of PPO ensures agents learn to develop policies that will be subsequently used in the testing phase. Analysis of the trained models is conducted through the following experiments. These include the length of time agents train for (specified in time steps) and the stimuli presented to prey agents during the training phase (in this case, the presence or absence of the predator agent) and, in turn, the impact of this stimulus on their subsequent behaviour in the testing phase.

In the first experiment, identical initial populations of prey agents are compared across two conditions. In condition one, agents are trained for 580,000 cycles (Table \ref{tab:exp1_summary}, Model Condition one). In the second condition, they are trained for 1,000,000 cycles (Table \ref{tab:exp1_summary}, Model Condition two). The hypothesis is that agents that train for longer may develop more effective strategies which they utilise within the test scenario. A task-efficiency measure is conceived to evaluate agents under both configurations to assess if this is the case. This is achieved by measuring the amount of reward and penalties collected by agents under each configuration. A single dependent variable was produced by combining these measures, consisting of a formula containing the mean of the total positive points collected $PosTotal$. The mean of the total negative points $NegTotal$. Finally, the mean of the total number of times agents are caught $CaughtTotal$ which produces the task efficiency formula:

\begin{equation}\label{5.1}
(PosTotal \times 1) + (NegTotal \times -0.2) + (CaughtTotal \times -1)
\end{equation}

The Task Efficiency formula is weighted by reward and penalty values. The rewards and penalties, set during the training parameter selection stage, allow prey agents to know that getting caught by the predator and foraging negative points lead to negative outcomes while foraging positive point objects lead to positive outcomes.
The rewards for foraging positive points are +1, and the penalty for foraging a negative point is -0.2, while the penalty for being caught by the predator is -1. The task efficiency formula was devised to distinguish between agents that forage positive points while avoiding negative points and the predator. Compared to agents with lower task efficiency, where agents were less capable in foraging positive points and avoiding penalties. In both experiments, rewards and penalties averaged over multiple model runs over two conditions are compared. These are short training and long training for experiment one and the predator's presence either pre or post-training or both for experiment two. This research is interested in the behaviours that emerge when prey agents interact with a predator; however, some extra elements such as points are added to the environment to introduce spatial complexity. This ensures prey agents train to achieve a goal such as foraging positive points (a metric used to identify how well they are doing) and not randomly roaming the environment waiting to encounter the predator. The penalty for foraging a negative point is smaller than being caught by the predator as we wish to introduce complexity toward agents' decision-making process. Similarly, if the penalty for foraging a negative point was -1, then the severity of foraging negative points and being caught by the predator would be equal. Consequently, prey agents may prefer being caught by the predator in certain situations.

In the second experiment, interest is centred around the notion of behavioural adaptation to stimuli. This idea is based on how RL agents behave when presented with stimuli in the testing phase, which were not present during the training phase (some sub-optimal measure of their adaptability). Furthermore, how this impacts agent decision-making relative to other agents exposed to the stimuli during training is quantified. As a result, these agents should have already developed behaviours to respond. Three model configurations are compared across the previously envisaged task-efficiency measure (Formula \ref{5.1}) to explore this. Using identical initial populations of prey agents, in the first model condition, the effectiveness of prey agents who train without the predator agent and complete their task in the testing phase without the predator is measured (Table \ref{tab:exp2_summary}, Model Condition one). This baseline experiment provides a comparative measure of the upper bounds of task efficiency in our model. In the second model condition, task efficiency for prey agents who train with the predator and subsequently test with the predator present is measured (Table \ref{tab:exp2_summary}, Model Condition two). In the final condition, the task efficiency of prey agents who train without the predator but are tested with the predator is observed (Table \ref{tab:exp2_summary}, Model Condition three).

Due to the stochastic nature of each post-training test, to verify the experiments, each model condition is tested fifty times for the same duration across all conditions. A summary of results, including task efficiency, can be viewed in the following sub-sections.

\subsection{Experiment 1, exploring the impact of training length on task efficiency}
\label{experiment1_subsection}
Experiment one was developed to identify the impact training length has on how well agents complete tasks. Are agents that adopt RL as a decision-making process, if trained for longer, more effective at their task?

\begin{table}[!h]
    \centering
    \resizebox{12cm}{!}{\begin{tabular}{p{4cm}p{3cm}p{3cm}}
    \toprule
    Model Condition & 1 & 2
    \\\midrule
    Training Cycles & 580k & 1 mil
    \\
    Predator in Training & Present & Present
    \\
    Predator in Testing & Present & Present
    \\
    Positive Points & 326.88 (34.231) & 779.4 (57.434)
    \\
    Negative Points & 320.44 (35.595) & 706.88 (53.059)
    \\
    Caught by Predator & 55.48 (15.931) & 72.7 (11.784)
    \\
    Task Efficiency & 207.312 (33.835) & 565.324 (53.164)
    \\\bottomrule
    \end{tabular}}
    \caption{Summary of the mean and (std) for each variable including task efficiency measure over all experiment one model conditions.}
    \label{tab:exp1_summary}    
\end{table}

When inspecting model conditions one and two (Table \ref{tab:exp1_summary}), it becomes clear that training agents for an extended period lead to agents that can learn better policies such as foraging more positive points. The task efficiency for model condition two has done relatively better than model condition one. However, agents in model condition two still forage a relatively large amount of negative points. Furthermore, agents are caught more often compared to model condition one; this shows that positive point foraging in model condition two outweighs the penalties for negative points and being caught by the predator. Due to the weighting of Formula \ref{5.1}, we would expect to see Caught\_by\_Predator occur less often than Negative\_Points, while Positive\_Points would be greater than the former two variables.

Statistics accompanying the quantitative results from Table \ref{tab:exp1_summary}, can be found in Table \ref{tab:exp1_ANOVA}.

\begin{table}[!h]
    \centering
    \resizebox{12cm}{!}{\begin{tabular}{p{5cm}p{2cm}p{2cm}p{2cm}p{2cm}p{2cm}}
    \toprule
    Variable & Condition 1 & Condition 2 & F score & p & Cohen's d
    \\\midrule
    Task Efficiency ($Pos - Neg - Caught$) & 207.312 & 565.324 & 1613.742 & 0.000 & -8.034
    \\
    Positive Point & 326.88 & 779.4 & 2290.235 & 0.000 & -9.571
    \\
    Negative Point & 320.44 & 706.88 & 1829.039 & 0.000 & -8.553
    \\
    Caught by Predator & 55.48 & 72.7 & 37.758 & 0.000 & -1.228
    \\\bottomrule
    \end{tabular}}
    \caption{Summary One-Way ANOVA and Cohen's d results over all experiment one model conditions.}
    \label{tab:exp1_ANOVA}    
\end{table}

It should be noted that the p values for Tables \ref{tab:exp1_ANOVA} and \ref{tab:exp2_ANOVA} should be read with caution, as more observations will make the One-Way ANOVA significant.

The outcome of the statistical tests (Table \ref{tab:exp1_ANOVA}) shows that the data collected from these two model conditions vary. As demonstrated by Cohen's d, the difference between the means of the task efficiency for the two groups is large; this confirms the earlier point that agents that train for longer are better at foraging positive points than agents who train for a shorter time. However, the same cannot be said for avoiding the predator. To conclude, agents that train for longer are better at finding optimal solutions for foraging but lack the same level of strategic behaviour to avoid the predator than agents trained for a shorter time.

\subsection{Experiment 2, exploring the impact of stimuli on task efficiency}
\label{experiment2_subsection}
Experiment two was conceived to determine how agents adapt to the presence of an unknown stimulus. For this example, the stimulus is the predator agent.

In this experiment, three model conditions are compared, the independent variables are; the presence of the predator either in training, testing or both. To examine how well agents perform relative to the predator, the task efficiency measure is used.

When comparing task efficiency across these conditions, several things become apparent. In model conditions where the predator is not present during training (Table \ref{tab:exp2_summary}, Model Conditions one and three), the difference between positive and negative points is large. Agents tend to collect more positive points while keeping the negative points minimal. However, agents are caught more often in model condition three compared to model condition two, which suggests agents that have not learnt policies to deal with the predator are caught more often.

\begin{table}[!h]
    \centering
    \resizebox{12cm}{!}{\begin{tabular}{p{3cm}p{3cm}p{3cm}p{3cm}}
    \\\toprule
    Model Condition & 1 & 2 & 3
    \\\midrule
    Training Cycles & 1 mil & 1 mil & 1 mil
    \\
    Predator in Training & Not Present & Present & Not Present
    \\
    Predator in Testing & Not Present & Present & Present
    \\
    Positive Points & 1455.18 (111.235) & 779.4 (57.434) & 1476.62 (122.026)
    \\
    Negative Points & 491.24 (42.519) & 706.88 (53.059) & 504.98 (44.173)
    \\
    Caught by Predator & 0 & 72.7 (11.784) & 102.08 (14.475)
    \\
    Task Efficiency & 1356.93 (108.803) & 565.324 (53.164) & 1273.544 (124.072)
    \\\bottomrule
    \end{tabular}}
    \caption{Summary of the mean and (std) for each variable including task efficiency measure over all experiment two model conditions.}
    \label{tab:exp2_summary}    
\end{table}

Given all three model conditions, when agents train without the predator, they collect more positive points than agents that train with the predator present. Furthermore, agents trained with the predator weigh the risks of getting caught with the risk of foraging a negative point (Table \ref{tab:exp2_summary}).

Comparing the task efficiency of model conditions two and three (see Table \ref{tab:exp2_ANOVA}) shows agents in model condition three produce statistically significantly higher task efficiency scores than those in model condition two. These results may seem counter-intuitive. However, one way to interpret this outcome is that agents in model condition two have likely devised policies that encourage predator avoidance to the detriment of foraging positive points. Conversely, in model condition three, agents are focused solely on foraging positive points. As emphasised by Cohen's d, the effect size of model condition two compared to model conditions one and three is large; this shows the predator significantly impacts how well prey agents forage.

\begin{table}[!h]
    \centering
    \resizebox{12cm}{!}{\begin{tabular}{p{5cm}p{2cm}p{2cm}p{2cm}p{2cm}p{2cm}}
    \toprule
    Task Efficiency ($Pos - Neg - Caught$) & Condition 1 & Condition 2 & F score & p & Cohen's d
    \\\midrule
    Model condition 1 vs 3 & 1356.93 & 1273.544 & 12.767 & 0.001 & 0.714
    \\
    Model condition 2 vs 3 & 565.324 & 1273.544 & 1376.41 & 0.000 & -7.420
    \\
    Model condition 1 vs 2 & 1356.93 & 565.324 & 2136.585 & 0.000 & 9.244
    \\\bottomrule
    \end{tabular}}
    \caption{Summary of One-Way ANOVA and Cohen's d results for Task Efficiency over all experiment two model conditions.}
    \label{tab:exp2_ANOVA}    
\end{table}

Collectively the results of both experiments indicate that within the simulation:
\begin{itemize}
    \item Agents that train for longer are more effective in devising goal-oriented strategies (experiment one model condition two, experiment two model conditions one, two and three).
    \item Agents that weigh the risks between multiple penalties (negative points and being caught by the predator) perform sub-optimally in achieving a goal (experiment one, both model conditions and experiment two, model condition two) compared to agents that focus solely on a single reward and penalty (experiment two, model condition one).
\end{itemize}

Agents that train with the predator present weigh the risks between rewards and penalties. This cannot be said with certainty for agents that are trained without the predator present. The behaviours that emerge from agents in both experiments (experiment one, both conditions, experiment two conditions one and three) should differ. We expect to observe more sophisticated behaviours from experiment one as agents devise different ways to avoid the predator while acquiring positive points compared to experiment two model conditions one and three where agents ignore the predator.

The quantitative results above provide a means to make confident assertions regarding how well prey agents have done foraging and avoiding the predator. The spatio-temporal patterns that emerge from these behaviours must not be neglected. It would be helpful to compare the occupied spaces within the environment for both agent types against the different model conditions. These data should allow us to observe the effect the predator has on prey agents.

\begin{figure}[!h]
   {\captionsetup{position=bottom,justification=centering}
    \begin{subfigure}[b]{0.5\textwidth}
        \includegraphics[width=\textwidth]{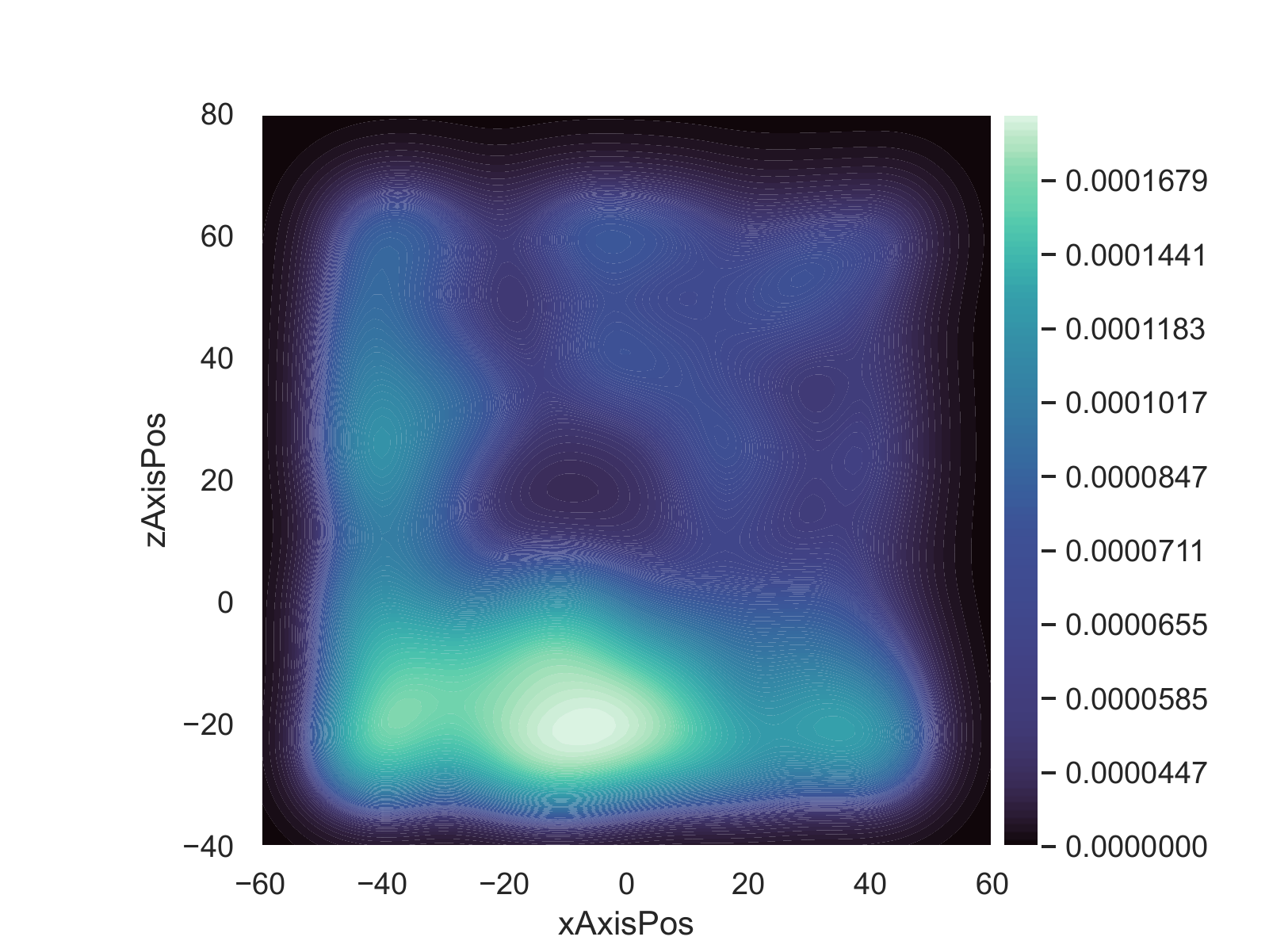}
        \caption{Prey: Experiment 1, Model Condition 1.}
        \label{fig:prey_exp1_mc1}
    \end{subfigure}
    ~
    \begin{subfigure}[b]{0.5\textwidth}
        \includegraphics[width=\textwidth]{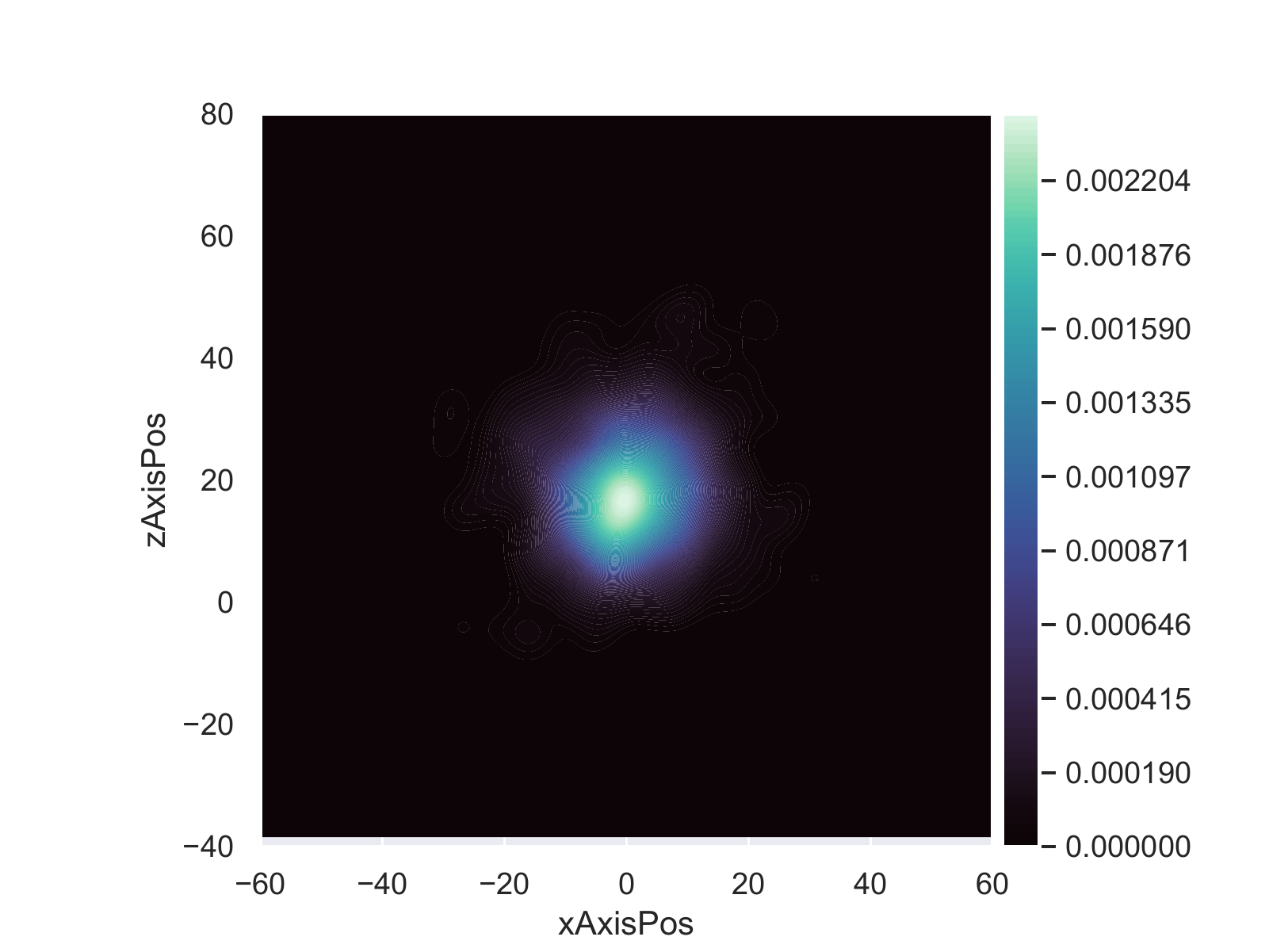}
        \caption{Predator: Experiment 1, Model Condition 1.}
        \label{fig:predator_exp1_mc1}
    \end{subfigure}
\\\vspace{6pt}
    \begin{subfigure}[b]{0.5\textwidth}
        \includegraphics[width=\textwidth]{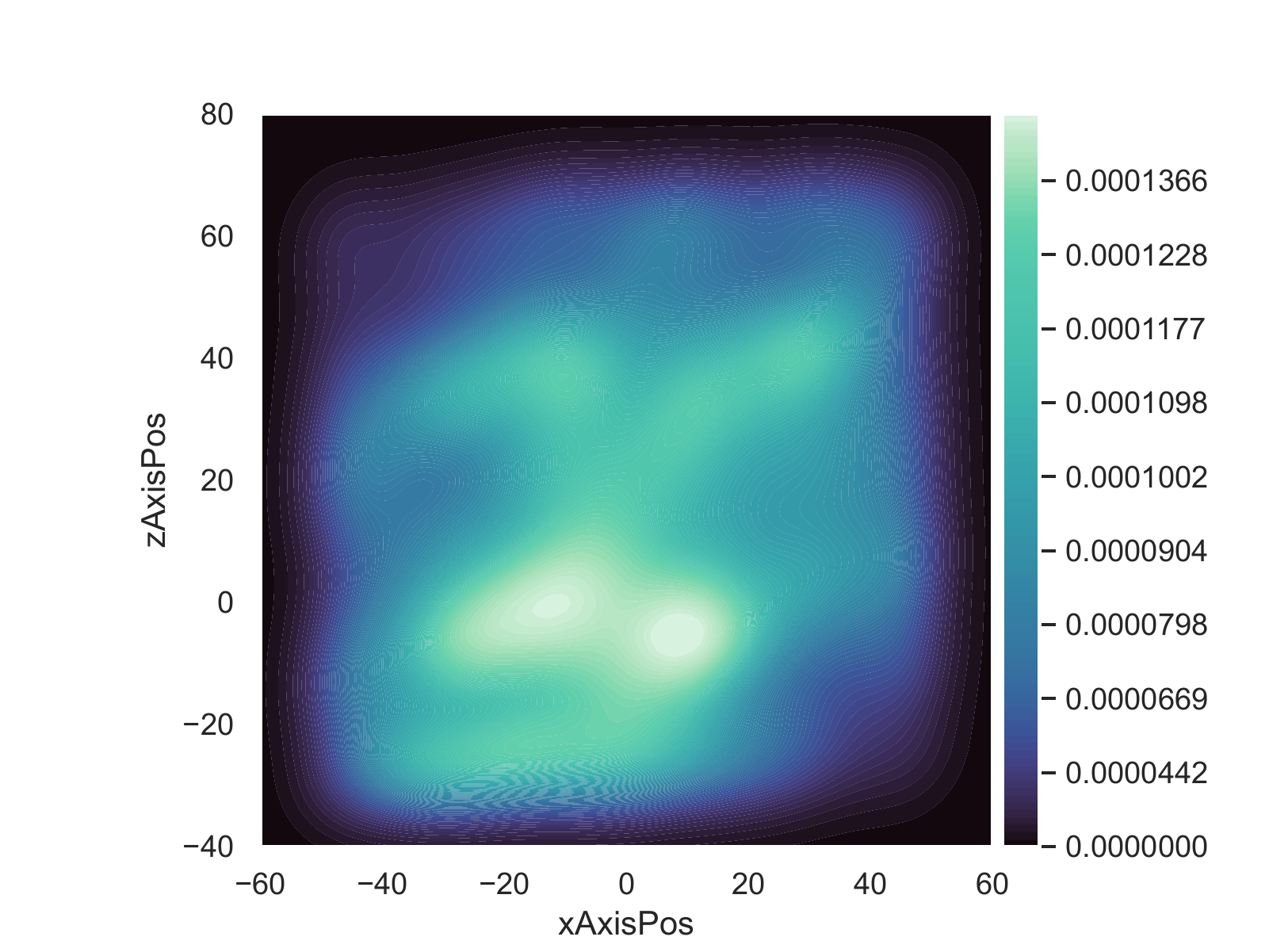}
        \caption{Prey: Experiment 1, Model Condition 2.}
        \label{fig:prey_exp1_mc2}
    \end{subfigure}
    ~
    \begin{subfigure}[b]{0.5\textwidth}
        \includegraphics[width=\textwidth]{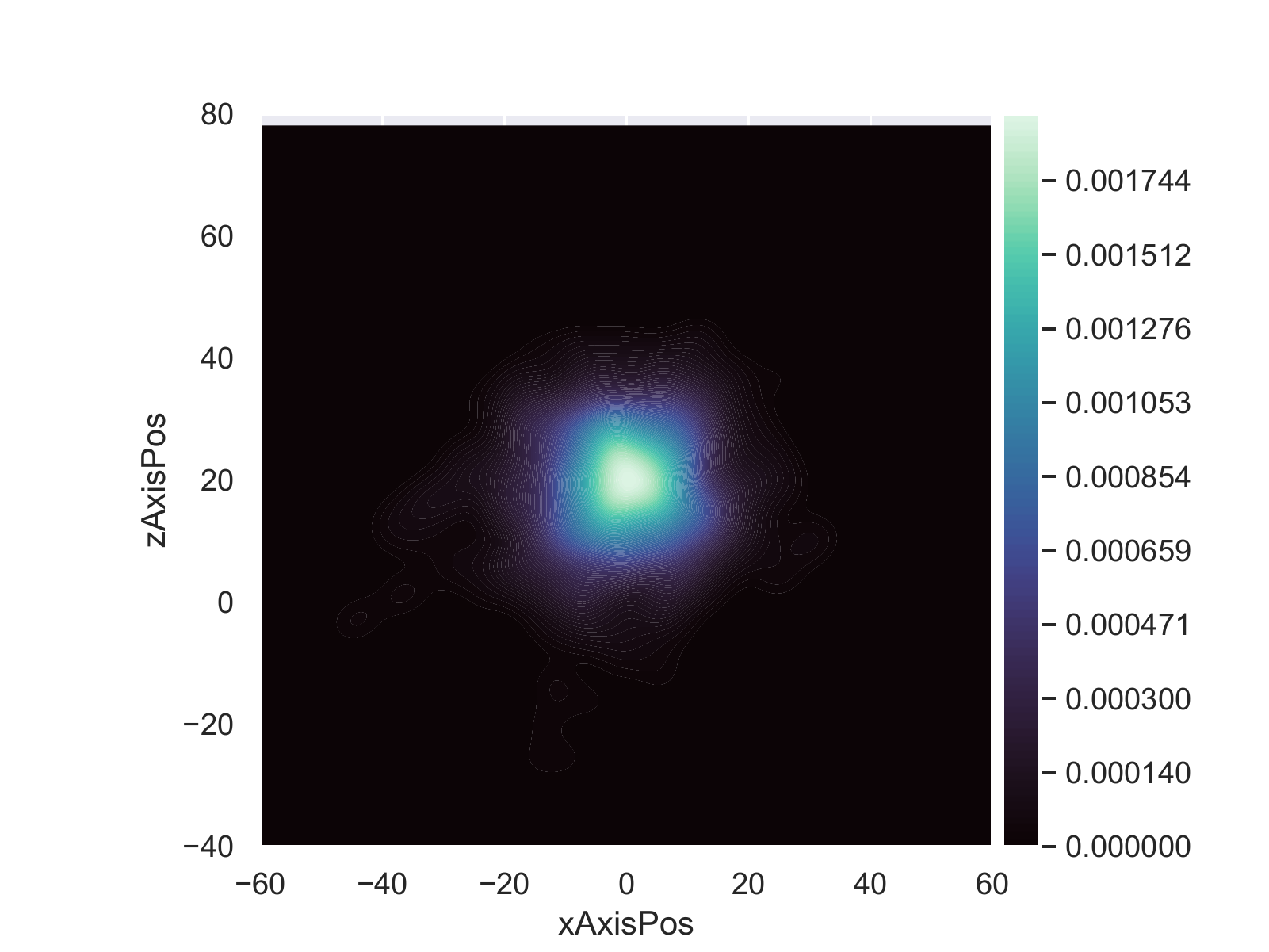}
        \caption{Predator: Experiment 1, Model Condition 2.}
        \label{fig:predator_exp1_mc2}
    \end{subfigure}}
\\\vspace{6pt}
    \begin{subfigure}[b]{0.5\textwidth}
        \includegraphics[width=\textwidth]{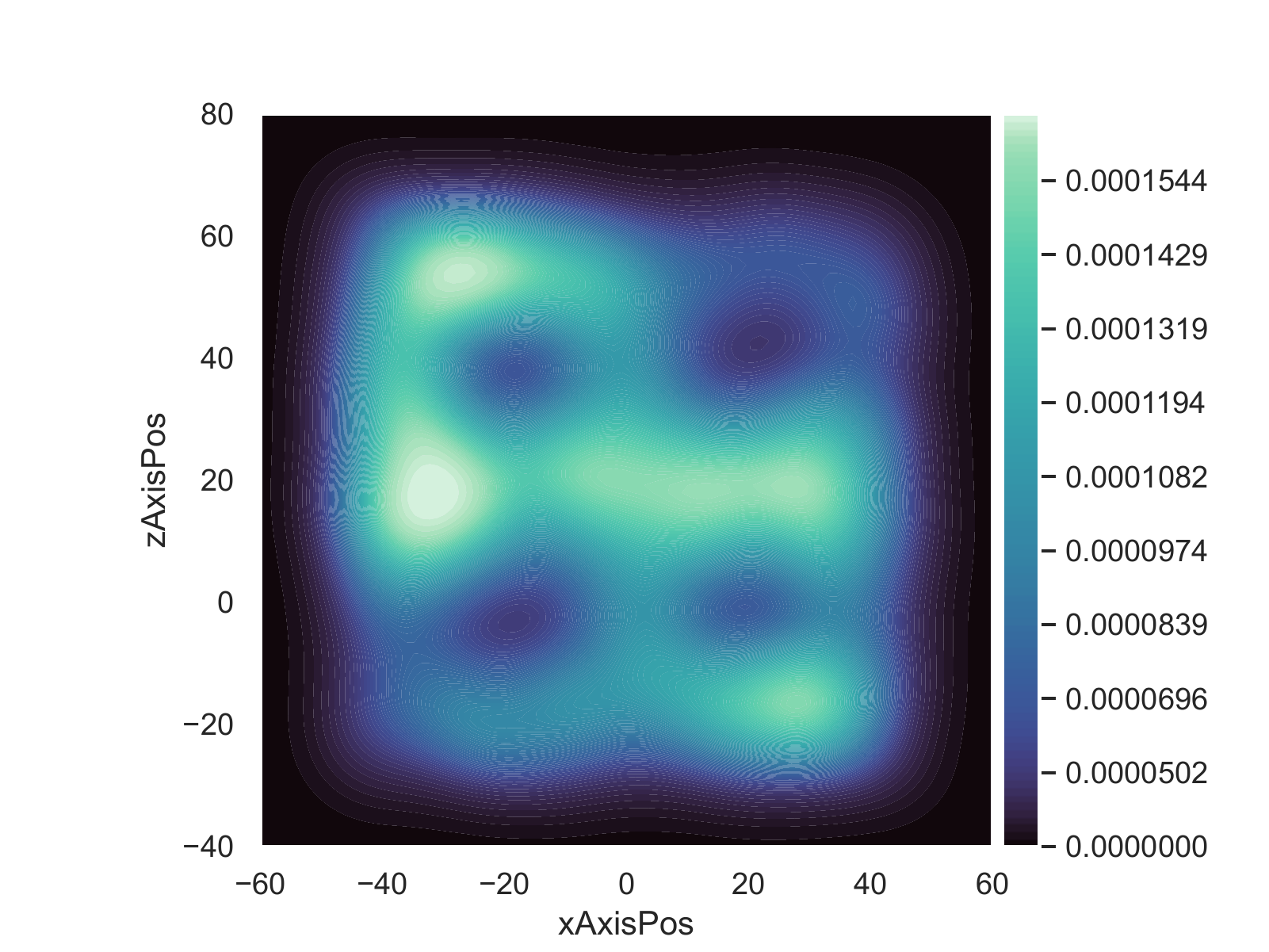}
        \caption{Prey: Experiment 2, Model Condition 3.}
        \label{fig:prey_exp2_mc3}
    \end{subfigure}
    ~
    \begin{subfigure}[b]{0.5\textwidth}
        \includegraphics[width=\textwidth]{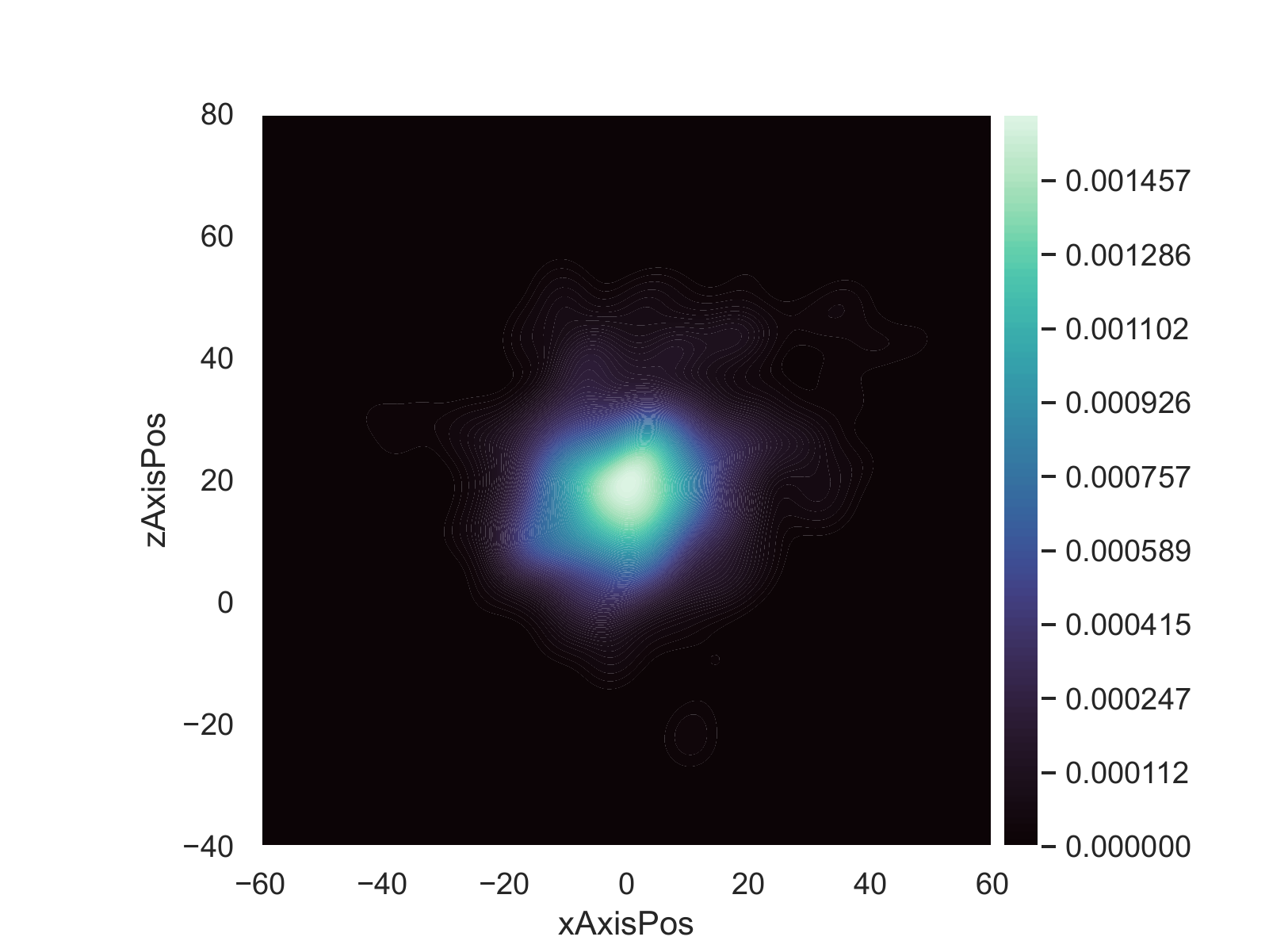}
        \caption{Predator: Experiment 2, Model Condition 3.}
        \label{fig:predator_exp2_mc3}
    \end{subfigure}
    \caption{A sample of spatio-temporal KDE plots of prey and predator movement patterns.}\label{fig:spatio-temporal-patterns}
\end{figure}

The Spatio-temporal movement patterns re-enforce the findings from Table \ref{tab:exp1_summary} showing how well the prey agents do in avoiding the predator for model condition one (Figures \ref{fig:prey_exp1_mc1} and \ref{fig:predator_exp1_mc1}), compared to model condition two (Figures \ref{fig:prey_exp1_mc2} and \ref{fig:predator_exp1_mc2}) in experiment one. The prey agents in experiment one, model condition one learn to avoid the centre of the environment where the predator occupies; for model condition two, we do not see similar patterns; the prey agents move diagonally more often and explore most of the environment. This is possibly the consequence of training for longer; thus, learning for longer makes the prey agents more goal-oriented (forage points) to the detriment of avoiding the predator (Figures \ref{fig:prey_exp1_mc2} and \ref{fig:predator_exp1_mc2}). In experiment two, model condition three (Figures \ref{fig:prey_exp2_mc3} and \ref{fig:predator_exp2_mc3}), the prey agents explore the entire environment while avoiding the barriers. Prey do navigate in areas occupied by the predator and get caught more often; however, compared to scenarios where prey are trained with the predator, agents are caught less often (Figures \ref{fig:prey_exp1_mc2} and \ref{fig:predator_exp1_mc2}). To conclude, when agents are trained using RL under various conditions, the spatio-temporal patterns reflect these differences. These quantitative results show that emerging complex behaviours are likely to be observed at an individual level; we see that the two experiments, namely, training length and unknown stimulus, lead to different spatio-temporal patterns. 

To answer the main research question, the agents' behaviours for these experiments must be analysed. Agents are more effective in avoiding the predator for some experiments. Conversely, in other experiments, agents perform more strongly in foraging positive points. To better understand these differences at an individual level, the behaviours of agents at an individual level need to be examined.

\newpage
\subsection{Individual Behaviour Analysis}
This sub-section describes the behaviours traced given the previous experiments conducted to explore whether intelligent adaptive behaviours occur when agents operate under the PPO framework. To recap, agents trained with the predator weigh the risks of getting caught with foraging a negative point. Furthermore, agents that train for longer outperform agents that train for a shorter time. Lastly, agents trained without the predator focus solely on foraging points; thus, they outperform their peers trained with the predator. However, these agents are caught by the predator more often.

This sub-section contains behaviours traced from all model conditions, and they are described frame-by-frame following the systematic approach below;

\begin{enumerate}
    \item Recording the experiment from start to finish.
    \item Playing back the experiment recording and taking note of the model scenario, i.e. one, two or three.
    \item Watching the movement of the predator agent to see if it interacts with a prey agent or if the predator is not present, only focusing on the prey.
    \item Every time the predator interacts with the prey; the video is paused and played frame-by-frame.
    \item The behaviour of the prey is captured frame-by-frame during the encounter until it has moved away and continues foraging points (known as the foraging behaviour).
    \item The frame-by-frame interactions are inspected closely; the behaviours observed are noted and compared with the quantitative analysis from the experiment sub-sections \ref{experiment1_subsection} and \ref{experiment2_subsection} to try to interpret how the prey agent has behaved.
\end{enumerate}

At the current time, the only way learned, intelligent behaviours can be identified is by visualising the model runs and qualitatively interpreting these behaviours; therefore, this was the chosen methodology. An accompanying video of some behaviours can be found at the following link: \url{https://youtu.be/-0bozJWC6l4}.

The visually observed behaviours from the model are utilised, as these are the best indicators of intelligent adaptive behaviours in the model, as has been the case in past literature \cite{Juliani2018Unity:Agents, OLSEN2015118, Jalalimanesh2017Simulation-basedLearning, Lopes2018IntelligentLearning, Spatharis2019CollaborativeManagement, 6151508}.

Following the visual inspection approach, several important examples of behaviours traced during each experiment are described below. These results come in the form of figures; each frame in a figure represents a state of the model from 1 to N time steps. The \textbf{red box} in each figure focuses the viewer on where the specific behaviour in question is occurring. The \textbf{green circle} indicates a prey agent and the \textbf{orange circle} is the predator agent.

\subsubsection{Hiding Behaviour}
\begin{figure}[!h]
        \centering
        \includegraphics[width=1\textwidth]{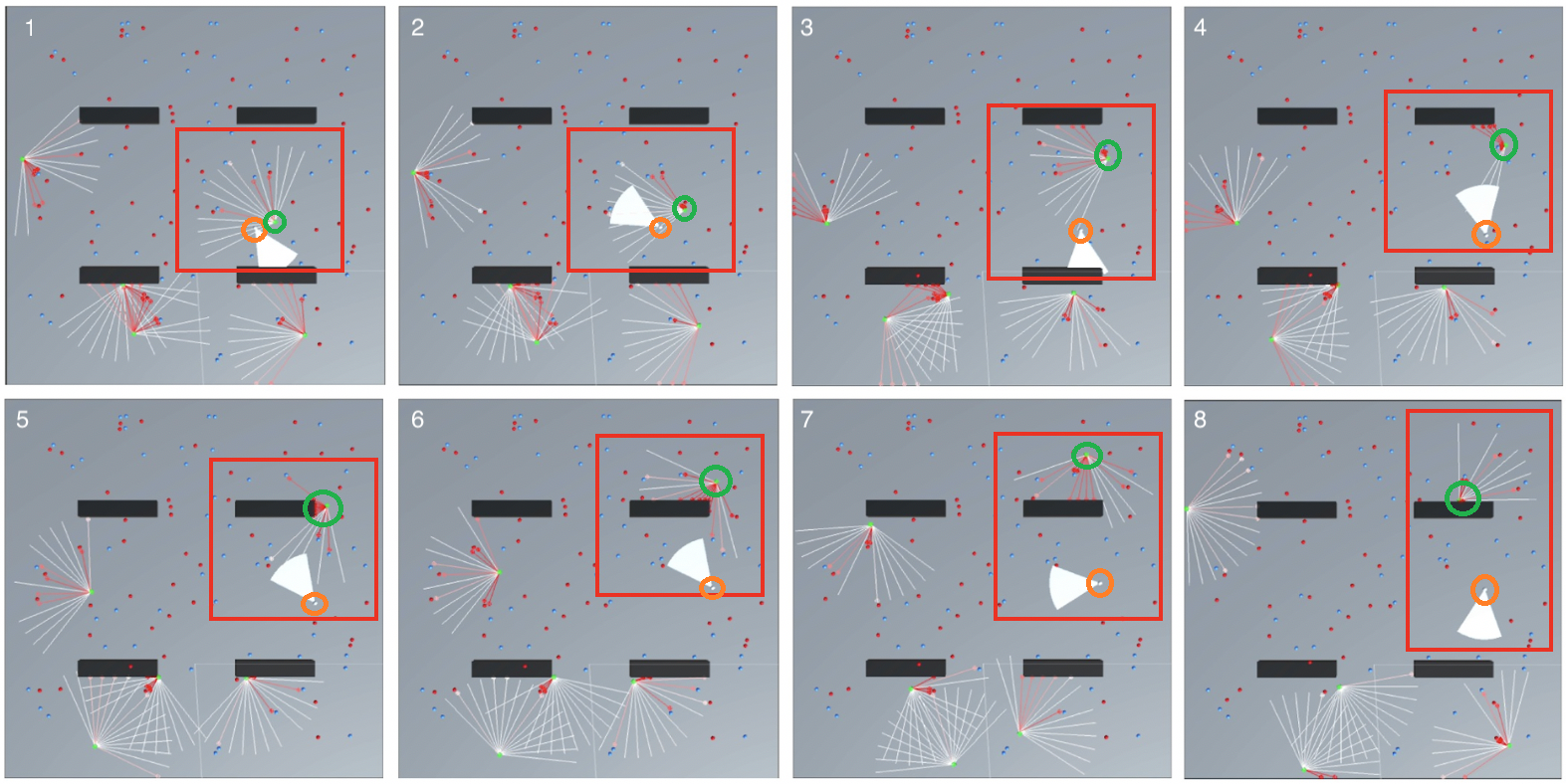}
        \caption{Experiment one, model condition one; a prey agent looking for a wall to hide behind.}
        \label{fig:behaviour_1.1}
\end{figure}

In Figure \ref{fig:behaviour_1.1}, a prey agent in the top right of the environment; spots the predator in the second frame; then it moves towards the opposite side of the closest barrier and hides; this behaviour indicates that prey agents have learned that the predator cannot see through obstacles in the environment. Consequently, the prey agent positions itself with its back towards the wall. This hiding behaviour is also observed in Figure \ref{fig:behaviour_1.3}. 

\subsubsection{Co-operative Behaviour}
\begin{figure}[!h]
        \centering
        \includegraphics[width=1\textwidth]{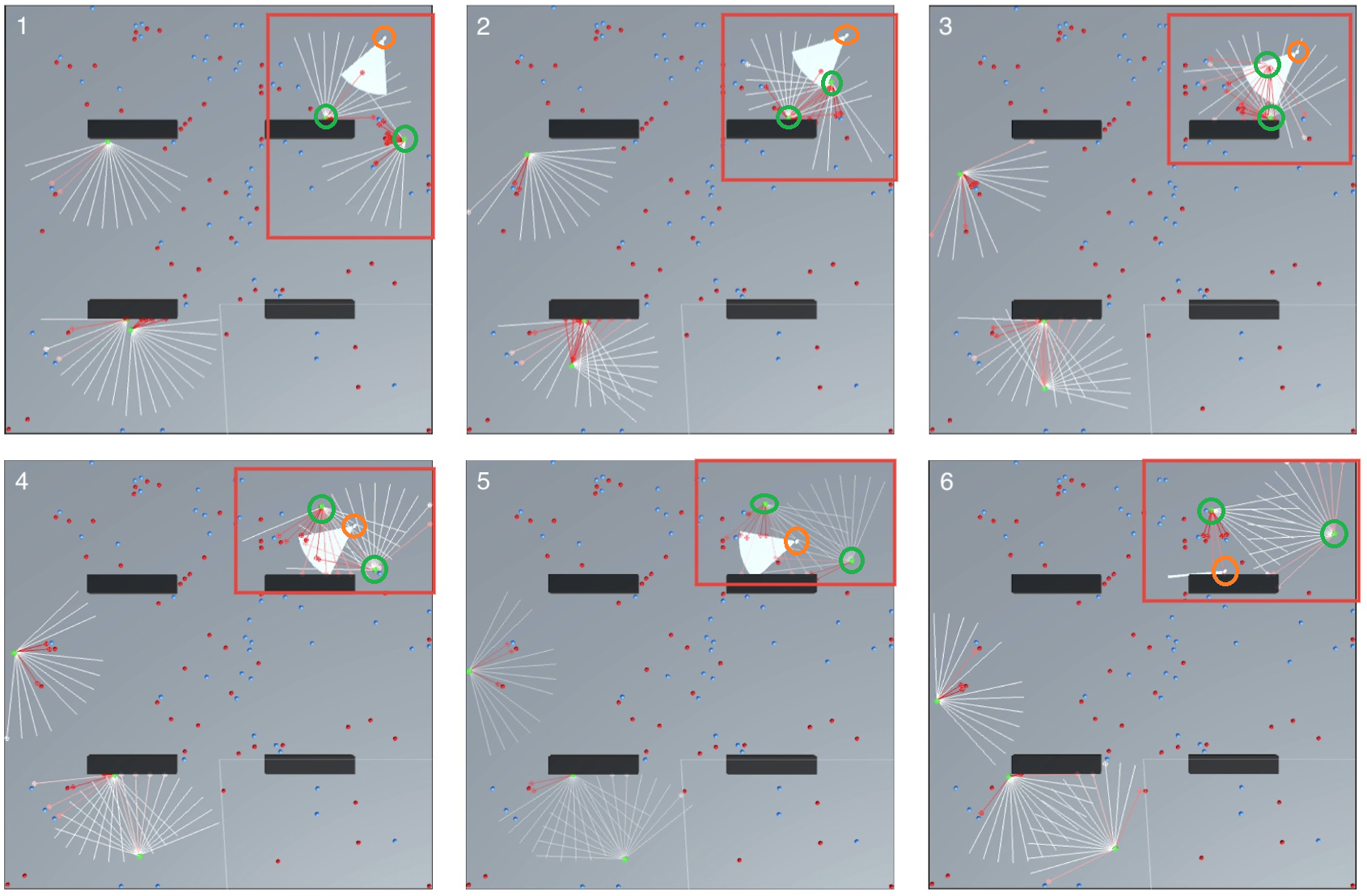}
        \caption{Experiment one, model condition one; two prey agents identify each other and move in opposite directions to avoid the incoming predator.}
        \label{fig:behaviour_1.2}
        \end{figure}

Prey agents are not designed to be adversarial nor cooperative; however, in experiment one, model condition one, what can be interpreted as cooperative behaviour is observed; however, this may be coincidental. Figure \ref{fig:behaviour_1.2} depicts a scenario from model condition one, where prey agents recognise the predator and move in opposite directions to evade capture. It could be argued that a wide range of actions could have led to a rewarding outcome. However, these prey agents adopt a policy that involves them moving away from each other.

\subsubsection{Evading Behaviour}
\begin{figure}[!h]
        \centering
        \includegraphics[width=1\textwidth]{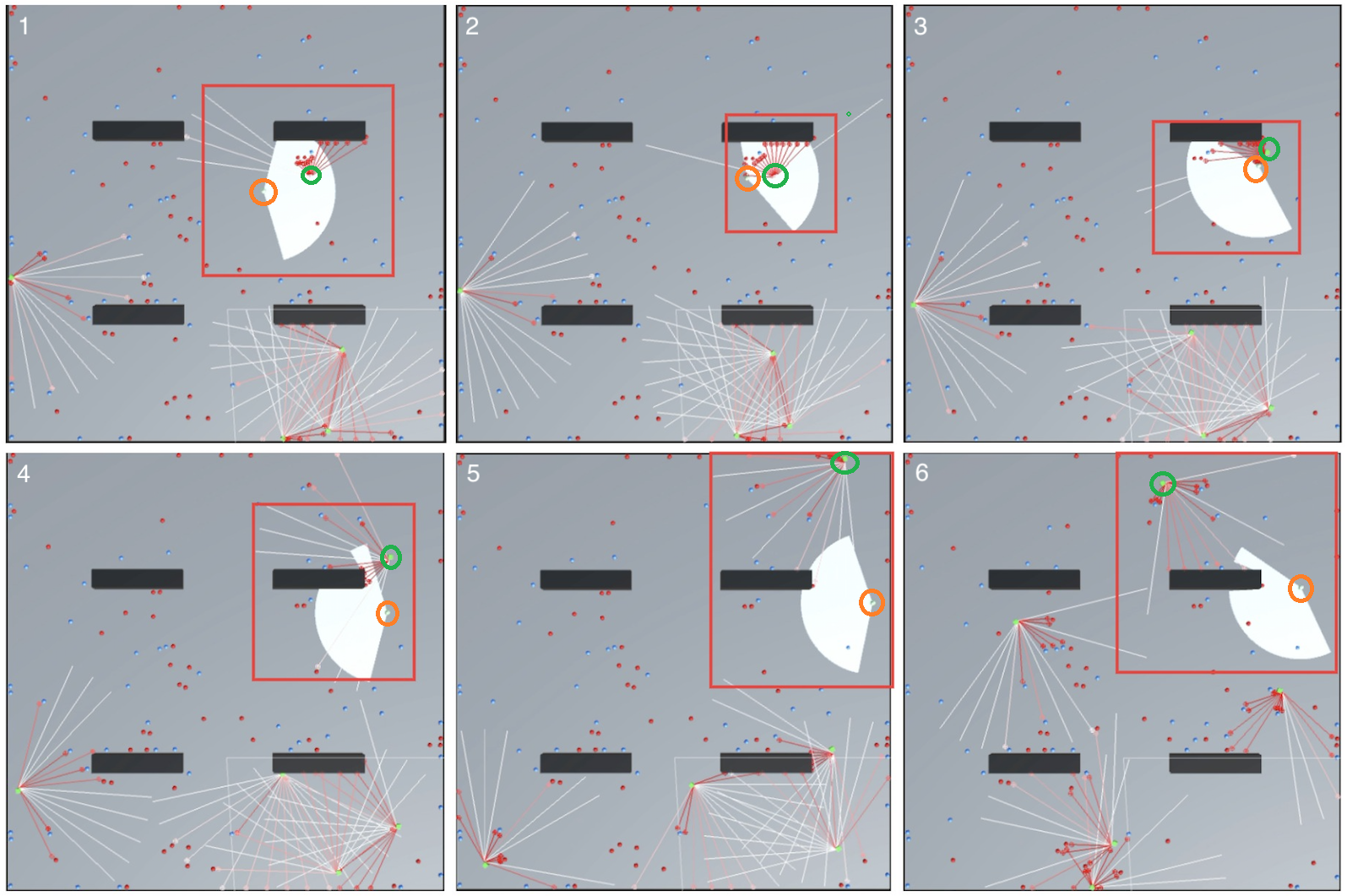}
        \caption{Experiment one, model condition two; a prey agent moves away from approaching predator and tries to use the barrier to evade the predator.}
        \label{fig:behaviour_1.3}
        \end{figure}

Figure \ref{fig:behaviour_1.3} depicts a blocking behaviour where a prey agent recognises a barrier, then realises the predator moving towards it; it successfully evades the predator and passes it using the barrier to block the predator's field of view. 

This learned behaviour considers the distance of the predator from the prey agent; as soon as the prey agent realises its presence, it takes immediate action to avoid it.  

\begin{figure}[!h]
        \centering
        \includegraphics[width=1\textwidth]{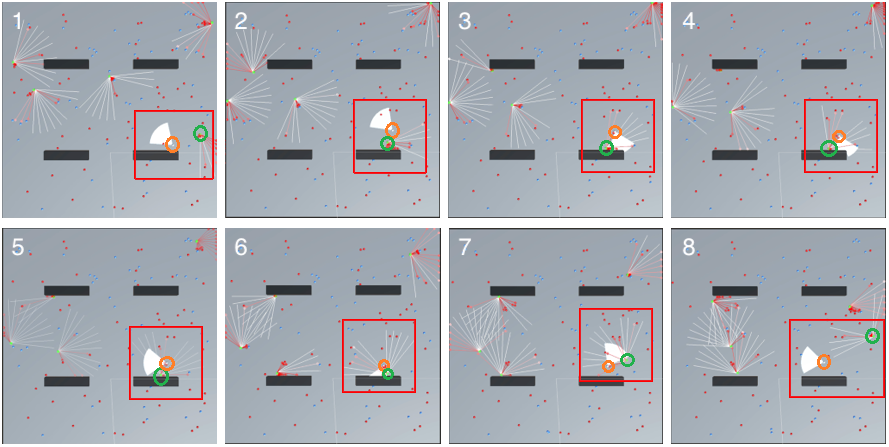}
        \caption{Experiment one, model condition two; a prey agent dodges incoming predator making it collide with the barrier.}
        \label{fig:Behaviour_2.1}
        \end{figure}

Figure \ref{fig:Behaviour_2.1} depicts a prey agent that intuitively dodges the incoming predator and allows it to collide with the barrier behind it. This behaviour might indicate prey agents have learned how quick predator agents move and thus can devise policies that take this information into account.

\subsubsection{Foraging behaviour}
\begin{figure}[!h]
        \centering
        \includegraphics[width=1\textwidth]{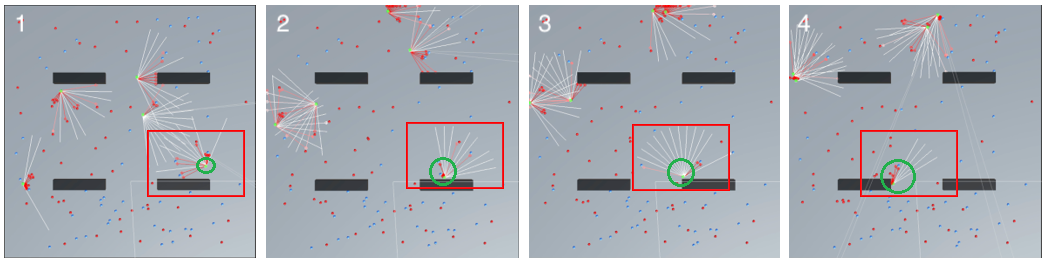}
        \caption{Experiment two, model condition one; the prey agents explore the environment foraging rewards.}
        \label{fig:Behaviour_3.1}
        \end{figure}

Figure \ref{fig:Behaviour_3.1} depicts foraging behaviour captured during experiment two, model condition one. In this experiment, we recall that prey agents developed policies in the absence of the predator agent. As the predator is not present, prey agents learn behaviours that only entail foraging points. 

\subsubsection{Circling Behaviour}
\begin{figure}[!h]
        \centering
        \includegraphics[width=1\textwidth]{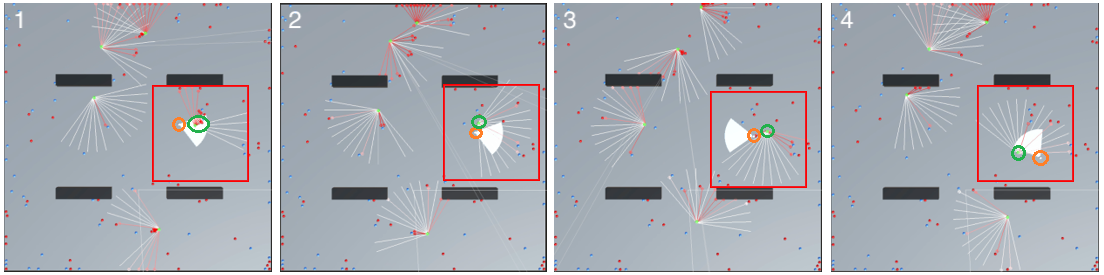}
        \caption{Experiment two, model condition three; prey agents move in a circular motion continuously while moving nearer to the closest positive point.}
        \label{fig:Behaviour_4.1}
        \end{figure}

In experiment two, model condition three (Table \ref{tab:exp2_summary}), prey agents are trained in a setting without a predator, then situated in an environment that contains a predator post-training. Agents appear to continue foraging positive points and avoiding negative ones. Agents move around in circles until they are close to a positive point; once they identify a positive point, they move to it (Figure \ref{fig:Behaviour_4.1}).

Prey agents are prone to being caught by the predator in experiment two, model condition three, as they have not devised strategies to deal with the predator (Table \ref{tab:exp2_summary}). However, prey agents have learned that moving in circles (Figure \ref{fig:Behaviour_4.1}) increases their chances of foraging a positive point while avoiding incoming negative points. This behaviour was not observed in any of the other conditions. This behaviour may have developed due to the sphere-shaped points moving on the environment's surface more often during training than previous model conditions.

This sub-section analysed the behaviours and outcomes of model conditions for both experiments using a systematic approach. Various behaviours were identified. These behaviours can be interpreted as `intelligent adaptive behaviours' such as hiding behind objects, evading the predator, using the environment to their advantage. The majority of these intelligent behaviours emerge from model conditions where the predator was present during training. These results show agents learn and apply policies that focus on foraging positive points while intuitively avoiding predators.

Agents trained without the predator focus on a specific task, i.e. forage points and avoid negative ones. There is a clear distinction between behaviours that emerge in conditions where the predator was present during training compared to conditions it was not. These results also show that agents trained using PPO can function in different situations/environments while achieving a goal.

\section{Discussion and Conclusion}
\label{section6}
This research set out to assess the effectiveness of developing an ABM in conjunction with the PPO technique as a behavioural framework. The research found that the RL technique supports agents in the behavioural evolution of intelligent decision making that resemble those observed in the real world. There have been several examples of ABMs that utilise RL. However,  the majority of these have either deployed RL as an extension to extract the optimal set of steps in decision making, or they have lacked any investigation of the impact of RL algorithm features, i.e. the impact of novel stimulus on the subsequent behaviours of agents \cite{Tellidou2006ADynamics, Thapa2005AgentCircumstances, Rahimiyan2010AnMarket, OLSEN2015118}. The model presented in this paper suggests that training time directly impacts an agent's ability to improve its behavioural goals, i.e. foraging a large number of rewards compared to agents that have trained for a shorter period. However, agents that devise policies influenced by multiple penalties weigh the impact of these penalties and try to minimise the most impactful compared to the less impactful. Furthermore, the research highlights the ability for agents to operate under conditions in which they were never trained to encounter and continue to perform relatively well. While the points mentioned above seem apparent, this research attempted to quantify the degree to which such outcomes transpired. Finally, where this research diverges from past literature is the fact that individual-level behaviours that were procured from the experiments were subjectively interpreted by following a set of comprehensive steps from procurement to interpretation.

Limitations of this work include the difficulty of identifying training parameters that allow the algorithm to train the model efficiently. If the parameter values are not appropriate for the model configuration this can negatively impact the outcome of training; for example, agents may behave erratically and diverge from achieving any goal. In this research, the default parameters provided by the software library ml-agents \cite{Juliani2018Unity:Agents} were adopted as these were tested extensively on multiple training environments; however, in future research, the impact of these choices should be assessed.

Another limitation of the research relates to the identification and interpretation of individual-level agent behaviours. Core to understanding the advantages and disadvantages of RL for developing realistic agent behaviours is assessing the behaviours generated by the RL algorithms. When writing this paper, the literature provides no agreed-upon methods for identifying and subsequently interpreting behaviours that agents enact during model testing. This limitation is an indicator of the lack of research done in this area. In response, this research attempted to identify a comprehensive set of steps in interpreting the observable behaviours that agents enact. However, this time-intensive technique may not be applicable for more complex models.

The accessibility of model development using novel technologies such as Unity and the subsequent programming language C\# is challenging. As the complexity of the ABM increases, so do the computational requirements. Due to rapid advancements in computing, this challenge is not insurmountable. 

Despite the lack of literature and steep learning curve in developing an ABM in Unity with RL, these techniques together can lead to valuable outcomes. The research shows that if PPO is adopted as a decision-making mechanism, agents can organically grow behaviours through achieving rewards and punishments. Furthermore, agents can weigh multiple risks and rewards and act accordingly. These attributes that agents develop can be observed in real-world situations, i.e. when people weigh the risks of being captured by the police before attempting a crime, or predatory animals weighing the risks of the prey escaping before deciding to pursue or ignore. This research indicates that agents can portray behaviours that would be considered "intelligent" without any explicit prior knowledge of these behaviours, which is one of the core strengths of RL. Moreover, agents have been shown to adapt to non-deterministic dynamic changes within the environment; these agents can continue achieving their relative goals within these scenarios.

An avenue for future research is using the techniques presented here to explore the relationships between people and enforced rules to prevent a contagious virus from moving through the population. This research is also applicable to population dynamics by simulating changing populations and the individual level interactions among populations.

To conclude, this research demonstrates that RL can provide a means for developing agents through training that exhibits 'intelligent' behaviours that evolve through space and time. This research shows that behaviours are encouraged by the environmental surroundings of agents. The experiments conducted highlight that training time impacts agent performance and that agents can adapt to environmental changes and behave sub-optimally when multiple penalties are considered. The spatial patterns of agent movement for each experiment condition vary, showing a strong spatial influence in decision making. The research demonstrates that agents with varying decision-making frameworks can co-exist within an environment, i.e. the predator agent applied simple if-then-else rules while prey agents utilised RL. This research shows that RL is a viable option as a decision-making framework for future ABMs and that the use of RL within ABM research could be revolutionary.

\section*{Data \& Model Access}
\begin{itemize}
    \item The model was developed in Unity using the C\# programming language. The predator-prey model and statistical analysis of simulation results can be found in the main ABM project's (Github) repository: \url{https://github.com/SedarOlmez94/Agent_Based_Modelling_Projects}
    \item Specifically, scripts and instructions to run the experiments are available at: \url{https://github.com/SedarOlmez94/Agent_Based_Modelling_Projects/tree/master/Predator-prey_RL_model}
\end{itemize}

\section*{Funding}
This project has recieved funding from the Economic and Social Research Council, grant number: ES/P000401/1; the Economic and Social Research Council and The Alan Turing Institute, grant number: ES/R007918/1.

\section*{Author Contributions}
SO - developed the agent-based model, including the reinforcement learning application, developed the experiments and analysis of the output data using data science techniques, wrote the paper and designed the study. DB - critically reviewed the article, suggested minor corrections. AH - critically reviewed the article, suggested minor corrections and acquired the funding for the project. All authors gave final approval for publication and agree to be held accountable for the work performed therein.

\section*{Notes}
\begin{enumerate}
    \item ML-agents Unity extension can be found here: \href{https://archive.vn/G1pSi}{here}.
    \item Unity can be downloaded \href{https://archive.vn/wL0Jn}{here}.
    \item ML-agents architecture documentation can be found \href{https://archive.vn/lgX8E}{here.}
    \item PPO hyper parameter best practices can be found \href{https://archive.vn/bILLG}{here}.
    \item ABM developed and trained on the following desktop: Intel Core i7-7700K, 32 GB RAM, 256 GB SSD Plus 2 TB HDD, 2 x NVIDIA GTX 1070 8 GB Graphics, Windows 10 Home.
    \item ABM is run on the following laptop: MacBook Pro (15-inch, 2018), 2.6 GHz Intel Core i7, 32 GB 2400 MHz DDR4, Intel UHD Graphics 630 1536 MB
\end{enumerate}

\newpage
\section*{Appendix A: Supplementary Materials}

\begin{table}[!h]
\centering
\resizebox{\textwidth}{!}{\begin{tabular}{p{3cm}p{11cm}p{2cm}}
\toprule
Variable & Description & Value\\
\midrule
Movement\_speed & The speed of movement & 20   \\
Rigidbody & The 3D agent object has a rigid body property \\
Capsule & The 3D agent's shape property \\
Capsule\_Collider & A component used to trigger an event when it comes into contact with another object\\
AIPredator.cs & The C\# script component allows the predator to perform actions in the environment \\
View\_camera & A camera component traces the movement of the agent in first person \\
Velocity & The velocity of the agent (take both axes X, Y normalise them then multiply by Movement\_speed) \\
View\_radius & The radius of the field of view for the agent & 10.33 \\
View\_angle & The angle of view within the View\_radius is between 0 and 360 & 80 \\
Target\_mask & A layer tag for objects the agent considers as targets & Target \\
Obstacle\_mask & A layer tag for objects the agent considers as obstacles & Obstacle \\
Visible\_targets & A list of all the targets the agent has seen \\
\bottomrule
\end{tabular}}
\caption{Predator agent's parameters.}
\label{tab:Predator_parameters} 
\end{table}

\begin{table}[!h]
    \centering
    \resizebox{\textwidth}{!}{\begin{tabular}{p{3cm}p{11cm}p{2cm}}
    \toprule
    Variable & Description & Value\\
    \midrule
    Rigidbody & The 3D agent object has a rigid body property \\
    Cube & The 3D agent's shape property \\
    Box\_Collider & A component used to trigger an event when it comes into contact with another object\\
    Prey.cs & The C\# script component allows the agent to perform actions in the environment \\
    Camera & A camera component traces the movement of the agent in first person \\
    Velocity & The velocity of the agent \\
    Turn\_speed & The speed of which the agent turns & 300 \\
    Move\_speed & The speed of which the agent moves on the X, Z axis & 2 \\
    Normal\_material & Normal material (agent is neither rewarded or penalised if it interacts with these) & GreenAgent \\
    Good\_material & Good material (agent is rewarded if it interacts with these) & PositivePoint \\
    Bad\_material & Bad material (agent is penalised if it interacts with these) & NegativePoint \\
    Use\_Vector\_Obs & If checked, the agent will send information to the neural network during training \\
    Ray\_Perception\_Sensor & A sensor which identifies objects within a given perimeter\\
    \bottomrule
    \end{tabular}}
    \caption{Prey agent's parameters.}
    \label{tab:Prey_parameters}    
\end{table}

\begin{table}[!h]
    \centering
    \resizebox{\textwidth}{!}{\begin{tabular}{p{2cm}p{4cm}p{8cm}}
    \toprule
    Component & Property & Value\\
    \midrule
    Plane & Position, Rotation and Scale along the X, Y, Z coordinates & P[0, 0 ,0], R[0, 0, 0], S[10, 1, 10] \\

    Wall 1 & Position, Rotation, Scale and Box Collider & P[-5.11, 0.8, 0], R[0, 0, 0], S[0.2, 1.7, 10], Box Collider [1, 1, 1]  \\
    Wall 2 & Position, Rotation, Scale and Box Collider & P[5.11, 0.8, 0], R[0, 0, 0], S[0.2, 1.7, 10], Box Collider [1, 1, 1]  \\
    Wall 3 & Position, Rotation, Scale and Box Collider & P[-0.04, 0.8, -5.07], R[0, 90, 0], S[0.2, 1.7, 10], Box Collider [1, 1, 1]  \\
    Wall 4 & Position, Rotation, Scale and Box Collider & P[-0.04, 0.8, 5.07], R[0, 90, 0], S[0.2, 1.7, 10], Box Collider [1, 1, 1]  \\
    Camera & Position, Rotation, Scale, Clear Flags, Culling Mask and Projection & P[-1.13409, 32.80403, 125.6395], R[26.565, -180, 0], S[1, 1, 1], Clear Flags = Skybox, Culling Mask = Everything, Projection = Perspective   \\
    Directional Light & Position, Rotation, Scale, Type and Mode & P[0, 3, 0], R[50, -30, 0], S[1, 1, 1], Type = Directional, Mode = Realtime \\
    \bottomrule 
    \end{tabular}}
    \caption{Environment parameters.}
    \label{tab:Environment_parameters}    
\end{table}

\begin{table}[!h]
\centering
\small
\resizebox{\textwidth}{!}{\begin{tabular}{p{3cm}p{15cm}p{2cm}}
\toprule
Parameter & Definition & Range\\
\midrule
Batch\_size & The amount of experiences that transpire in each iteration of gradient descent. This parameter must always be a fraction of buffer\_size. & (Continuous): [512, 5120], (Discrete): [32, 512]\\

$\beta$ & The strength of entropy regularisation. This parameter ensures that agents accurately explore the action space during training. Increasing this will ensure that more arbitrary actions are taken frequently. & [1e-4, 1e-2]\\

Buffer\_size & The quantity of observed sensory information (experiences) to accumulate before updating the policy model. This parameter corresponds to how many experiences (observations, actions and rewards obtained) should be secured before learning begins or updating the model. This should be a multiple of batch\_size. & [2048, 409600]\\

$\epsilon$ & This parameter determines how fast the policy can evolve during training. Epsilon corresponds to the adequate threshold of divergence between the old and new policies during gradient descent updating. Setting this value small will end in more stable updates but will also slow the training process. & [0.1, 0.3]\\

Hidden\_units & The number of units in the hidden layers of the neural network. & [32, 512]\\

GAE $\lambda$ & This parameter corresponds to the lambda parameter used when determining the Generalised Advantage Estimate (GAE). This can be considered how much the agent relies on its current value estimate when determining an updated value estimate. Low values correspond to relying more on the current value estimate (which can lead to bias), and large values correspond to relying more on the actual rewards received in the environment (which can be considerable variance). & [0.9, 0.95]\\

Learning\_rate & The initial learning rate for gradient descent. This parameter relates to the strength of each gradient descent update step. & [1e-5, 1e-3]\\

Max\_steps & The greatest number of simulation steps to run during a training session. & [5e5, 1e7]\\

Memory\_size & The size of the memory an agent must keep; this is generally utilised if use\_recurrent is true. The parameter relates to the size of the array of floating-point numbers used to store the hidden state of the recurrent neural network. This value must be a multiple of 4 and should scale with the amount of information the agent will need to remember to complete the task. & [64, 512]\\

Normalise & If true, will automatically normalise observations. This normalisation is based on the running average and variance of the vector observation. Normalisation can be effective in complex continuous control problems but may be harmful with more straightforward discrete control problems. & [true, false]\\

Num\_epoch & The number of passes to make through the experience buffer when performing gradient descent optimisation. Decreasing this will ensure more stable updates at the cost of slower learning. & [3, 10]\\

Num\_layers & The number of hidden layers in the neural network. This parameter corresponds to how many hidden layers are present after the observation input or after the Artificial Neural Network (ANN) encoding of the visual observation. For more minor problems, fewer layers are likely to train faster and more efficiently. More layers may be necessary for more significant control problems. & [1, 3]\\

Horizon (T) & Corresponds to the number of steps of experience to collect per-agent before appending it to the experience buffer. When this limit is reached before the end of an episode, a value estimate is used to predict the overall anticipated reward from the agent's current state. As such, this parameter trades off between a less biased but higher variance estimate (long time horizon) and more biased but less varied estimate (short time horizon). If there are many rewards within an episode or episodes are prohibitively large, a smaller number can be more ideal. & [32, 2048]\\

Sequence\_length & Corresponds to how long the sequences of encounters must be while training (only used with Recurrent Neural Networks). & [4, 128]\\

Summary\_freq & A Tensorboard specific parameter used to identify how often to log training statistics during a training session. & [1e3, 5e4]\\

Use\_recurrent & Set to true if a Recurrent Neural Network is to be used, else a default Artificial Neural Network (ANN) is applied. & [true, false]\\

$\gamma$ & This parameter corresponds to the discount factor for future rewards. This can be thought of as how remote into the future the agent should care about possible rewards. When the agent should be operating in the present to prepare for rewards in the distant future, this value should be large. In instances when rewards are more immediate, they can be smaller. & [0.8, 0.995]\\

\bottomrule
\end{tabular}}
\caption{Formal definition of the training parameters and recommended range, where [x, y] inclusive, source: \cite{Juliani2018Unity:Agents}.}
\label{tab:Hyper-parameters-formal-definition} 
\end{table}


\bibliographystyle{unsrt}
\clearpage
\bibliography{references}

\end{document}